\documentclass[9.5pt,journal,compsoc]{IEEEtran}
\usepackage{cite}
\usepackage{amsmath,amssymb,amsfonts}
\usepackage{algorithmic}
\usepackage{algorithm2e}
\usepackage[multiple]{footmisc}
\usepackage{graphicx}
\usepackage{textcomp}
\usepackage{hyperref}
\usepackage{colortbl}
\usepackage{enumitem}  
\usepackage{xcolor}
\usepackage{graphicx}
\usepackage{url}
\usepackage{amsthm} 
\usepackage[normalem]{ulem}
\usepackage{caption}
\usepackage{subcaption}

\DeclareMathOperator*{\argmax}{arg\,max}
\DeclareMathOperator*{\argmin}{arg\,min}
\newtheorem{theorem}{Theorem}

\newtheorem{lemma}{Lemma}
\newtheorem{proposition}{Proposition}

\usepackage[textsize=scriptsize]{todonotes}

\ifCLASSINFOpdf
\else
\fi
\begin{document}
%
\title{Optimal Influencer Marketing Campaign under\\ Budget Constraints using Frank-Wolfe}
%
%
%
%

\author{Ricardo L\'opez-Dawn,~and 
        Anastasios Giovanidis~\IEEEmembership{Member,~IEEE} 
\IEEEcompsocitemizethanks{\IEEEcompsocthanksitem The authors are with the Sorbonne Universit\'e, CNRS, LIP6, F-75005 Paris, France, 
\{Ricardo.Lopez-Dawn, Anastasios.Giovanidis\}@lip6.fr
\IEEEcompsocthanksitem An earlier version of this paper was presented at the 19th International Symposium on Modeling and Optimization in Mobile, Ad Hoc and Wireless Networks (WiOpt 2021). \IEEEcompsocthanksitem This work is funded by the ANR (French National Agency of Research) by the “FairEngine” project under grant ANR-19-CE25-0011.}
\thanks{Manuscript created March 31, 2022.}
}

%
%

\markboth{Journal of \LaTeX\ Class Files,~Vol.~14, No.~8, August~2015}%
{Shell \MakeLowercase{\textit{et al.}}: Bare Demo of IEEEtran.cls for Computer Society Journals}
%



\IEEEtitleabstractindextext{%
\begin{abstract}
Influencer marketing has become a thriving industry with a global market value expected to reach 15 billion dollars by 2022. The advertising problem that such agencies face is the following: given a monetary budget find a set of appropriate influencers that can create and publish posts of various types (e.g. text, image, video) for the promotion of a target product. The campaign's objective is to maximize across one or multiple online social platforms some impact metric of interest, e.g. number of impressions,  sales (ROI), or audience reach. In this work, we present an original continuous formulation of the budgeted influencer marketing problem as a convex program. We further propose an efficient iterative algorithm based on the Frank-Wolfe method, that converges to the global optimum and has low computational complexity. We also suggest a simpler near-optimal rule of thumb, which can perform well in many practical scenarios. We test our algorithm and the heuristic against several alternatives from the optimization literature as well as standard seed selection methods and validate the superior performance of Frank-Wolfe in execution time and memory, as well as its capability to scale well for problems with very large number (millions) of social users.

\end{abstract}

}

\maketitle

\IEEEdisplaynontitleabstractindextext

%
\IEEEpeerreviewmaketitle

\IEEEraisesectionheading{\section{Introduction}\label{sec:introduction}}

%
%
%
%
\IEEEPARstart{M}{arketing} 
 has a long history \cite{I1} and its practices have evolved over time adapting to changes in the available media. In the 20th century the predominant media was the magazine, then changed to radio, then to television, and more recently to the world-wide-web. In our age of Online Social Platforms (OSPs), marketing has evolved in such a way that any social user can play the role of an advertising channel. Nowadays, several brands make use of the social sphere to disseminate information, e.g. Puma \cite{I4}, which advertised their \#IgniteXT line among younger public through posts created by $61$ influencers. Another well-known example was Spotify with its \#thatsongwhen influencer campaign \cite{I3}, with the objective to increase its new subscribers. Many other examples are documented and can be found for various industries (Pepsi, H\&M, Dior, Dreamworks, etc.)


In 2021, there exist more than 1360 influencer marketing agencies, the average earned media value per \$1 spent has increased to \$5.78 \cite{I5,I6}, and it is estimated that the influencer industry is on track to be worth up to \$15 billion dollars by 2022. Due to the growing importance and magnitude of this market, a general framework is necessary to describe campaigns and propose algorithms to select influencers with the aim to maximize some advertising objective subject to the available monetary budget.

The objective of an advertising campaign is usually to maximize one (or more) of the following metrics \cite{I5} over one or several OSPs:
\begin{enumerate}
    \item \textit{Impressions}: The total number of times that any post related to the campaign has been displayed in the Newsfeeds of all users. 
    \item \textit{Reach}: The total number of different users that viewed one or more posts related to the campaign in their Newsfeeds. Impressions and reach are ways to quantify the spread of a campaign.
    \item \textit{Engagements}: These include the total number of likes, comments and re-posts related to the campaign. This metric captures the interactions received in an advertising campaign.
    \item \textit{Conversion/Sales}: Generally, these metrics quantify the ROI (Return-On-Investment), which measures the efficiency or profitability of the influencer recruitment and generated posts within the advertising campaign.
    
\end{enumerate}

Influencers can be divided into three categories based on their dissemination capacity:
\begin{itemize}
    \item \textit{Nano-influencers}: These possess small, niche, and highly engaged audience. Nano-influencers have the smallest number of followers, the highest engagement per post, biggest ROI, and they are easier to recruit.
    \item \textit{Micro-influencers}: These have the characteristics of being strongly connected with their audience, they tend to receive a lot of engagements per post and are cost-accessible to businesses of all sizes.
    \item \textit{Macro-influencers}: These usually have a very wide audience, which comes with a significantly higher cost per post than the micro-influencers, but also with a higher level of professionalism.
\end{itemize} 

The price per post of influencers varies depending on the type of the social media platform, the type of content of the post, the number of followers, the average number of engagements per post, the advertised product, etc. 

Given a monetary budget over a time period during which the campaign is deployed, a company will search for a basket of influencers to maximize its campaign objective (e.g. impressions, engagements, reach). It is important to note that most influencers charge per created post. They will normally not engage their total activity to the promotion of a single company/product, in order to preserve a personal style and offer post variety that keeps on feeding their followers' interest. 

\subsection{Related Literature} 

The first relevant papers about viral influencer marketing in OSPs are by Domingos et al. \cite{R1,R2} which introduced the influence maximization problem. Here the authors introduce the influencer's network value, i.e. the expected profit due to social propagation starting from this user. 
Kempe et al. in \cite{a1} studied the influence maximization problem as a discrete problem, with the following elements:
\begin{itemize}
\item A social graph with the users as the vertex set and social ties among the users as the edge set.
    \item A diffusion process describing how content is diffused among social neighbors over discrete steps.
\end{itemize}

In \cite{a1} the influencer selection problem is stated as follows: for a given natural number $k>0$, choose at most $k$ users of the social network called the seed set, such that the number of users influenced (reached) is maximized when the diffusion process is
completed. In research work that succeeded this, user costs and budgetary restrictions have been introduced to maximize various metrics such as, profit \cite{0K2}, influence \cite{a12}, revenue \cite{0K3}, or profit for coordinate campaigns in \cite{a7}. Other very relevant recent works consider a community-based approach for the influence maximization \cite{rec1}, the minimization of the impact of misinformation \cite{rec2}, budgeted influence maximization with tags in \cite{Nrec1}, the estimation of influence spread \cite{eise}, the use of community structure and node coverage \cite{oimp}, the combination with a spreaders' ranking algorithm \cite{imsr}, crowd emotion \cite{imbce}, the relation to echo chamber effect in \cite{Nrec2}, as well as the spread over social networks described as $k$-submodular function in \cite{0K1}. 
Most of these works result in NP-hard problems with sub-modular structure that can be sub-optimally solved in polynomial time using greedy approximation algorithms. 

However, the binary decision to include an influencer in the seed set or not does not necessarily model reality, because an influencer does not normally attribute his whole activity to the advertising campaign, rather a couple of posts. The user cost in reality is calculated per post or content produced rather than per recruited influencer.

Furthermore, the diffusion of information in the OSPs does not follow a specific model, and even if it does to a certain extent, this model is generally not available to the advertiser. To add more value to this argument, empirical evidence from experiments has shown that network heterogeneity, assortativity and susceptibility are important mechanisms shaping social influence \cite{nature}. However standard diffusion models such as the Linear Threshold and Independent Cascade used in \cite{a1} do not incorporate such features. and generally underestimate the spread of influence. However, taking a data-centric approach, information about the post impressions and engagements from online social platforms can be (and in practice are) collected by the advertising companies, hence there are data sets available that track the campaign results and the detailed influence from a social user to the others, without the need to assume anything about the underlying diffusion.

\subsection{Our Contribution}

In this work, we introduce a new formulation of the influencer selection problem. Our problem has continuous unknowns instead of discrete; it aims to find the optimal \textit{participation ratio} in the campaign for each user of the OSP in order to maximize the campaign objective under budget restrictions. The participation ratio per user is the proportion of generated posts in favour of the campaign for each user. This quantity will be zero for most users (depending on the budget), but those with non-zero participation can contribute a continuous proportion of their activity to the campaign. Our formulation takes advantage of the \textit{assumed known} user activity over a time period, the \textit{assumed known} cost per post of each influencer and the \textit{assumed availability} of collected data about Newsfeed impressions. 

\begin{table}[t!]
\centering
\resizebox{\columnwidth}{!}{
\begin{tabular}{||c c||} 
 \hline
 Influence Maximization \cite{a1} & Our Budgeted Portfolio Optimization \\ [0.5ex] 
 \hline\hline
 Discrete & Continuous \\
 Graph & User set \\ 
 Diffusion process &  Data set of Impressions \\
 Cost per user & Cost per post \\
 Multi-platform cannot be handle & Extensions can be handle\\[1ex] 
 \hline\hline
 Objective:   & Objective: \\ 
 \hline
 Maximize the number of &  \\
 users influenced when & Maximize the campaign objective \\
 the diffusion process is over & (Impressions, Conversion/Sales or Reach)\\ [1ex] 
 \hline\hline
 Return:  & Return: \\ 
 \hline
 Seed set &  Participation ratio per user\\
  \hline
\end{tabular}
}
\caption{Differences between approaches}
\label{table:1}
\end{table}

The main differences between our model and Kempe's approach are summarized in Table \ref{table:1}. To elaborate on the differences, in \cite{a1} the work concerns the spread of a single post, the knowledge of the diffusion process is necessary and the user selection is binary. On the other hand, in our model the spread of influence is achieved by posting over time, the knowledge of the number of impressions from each source to any other user Newsfeed should be known, and we search for a continuous rate per user.

The formulation of the influencer marketing optimization problem under budget constraints on a single OSP is provided in section \ref{sec:II}. In section \ref{sec:III}, we develop our projection-free algorithm for concave functions via first order methods based on the Frank-Wolfe iteration \cite{III12}, and as a corollary, we present a rule of thumb 
that could be practical for the design of any marketing policy. These fast low-complexity methods 
can be applied for very large network sizes encountered in real platforms. Three particular cases of campaign objective are treated in this work:

 \textit{Impressions/Engagements}: This case arrives when we consider the advertiser's campaign objective as linear. The optimal solution can be found by the simplex algorithm with computational complexity of order $\mathcal{O}(max((N-1) \log(N-1),D))$ where $N-1$ is the number of users excepting the advertiser 
 and $D \leq N^2$ is the total number of pairs of users (including the advertiser) who can possibly interact, in the sense that one creates content which appears in the other's Newsfeed. Hence the solution scales well with the number of users, when the social graph is sparse.

 \textit{Conversion/Sales}: Under the assumption that the purchasing propensity of users (or the ROI) varies depending on their exposure to product related content, we study campaign objectives that are concave with respect to user impressions i.e. functions that exhibit diminishing returns. To achieve near-optimal solution with guarantees, we propose the iterative algorithm based on Frank-Wolfe with complexity $\mathcal{O}(max((N-1) log(N-1),D))$ per iteration. In section \ref{sec:V} we work with the general $\alpha$-fairness utility family, 
a special case of which is proportional fairness having the sum of logarithmic functions as objective.

 \textit{Reach}: Another special case of $\alpha$-fairness is when $\alpha$ tends to infinity, which gives a Max-Min fairness solution. This specific utility maximizes the \textit{Reach} by maximising the number of selected influencers having non-zero participation in the campaign for a given budget. 

In section \ref{sec:IV}, our formulation is extended to include advertising over multiple social platforms and accounting for multiple types of content (text, image, video) each with a different cost per user. The performance evaluation of our algorithm and the rule of thumb as well as a sensitivity analysis is illustrated on synthetic networks and on a large Twitter data set in section \ref{sec:V}. Besides, our empirical results found in section \ref{sec:V} allow us to verify the results established by \cite{nature}.
Conclusions are drawn in section \ref{sec:VI} and the code is available on GitHub \cite{a43}.

\section{The 
Portfolio Optimization Problem}
\label{sec:II}

Let us first describe a generic social network platform,
such as Facebook, Twitter or Instagram. A set of users generate and share some content, denoted as posts, through the platform. Each user has a list of followers and a list of leaders. A user can simultaneously be follower and/or leader of others. As a follower, he (she) is interested in the content posted
by his (her) leaders. With each user a Newsfeed is associated, which is a list of received posts.

We consider a constant number $N$ of active users in a specific time window, forming
the set $\mathcal{N}$. Users are labelled by an index $n=1,...,N$. They can be nodes of a friendship graph, which we do not need to know. We denote by $\lambda^{(n)}$ [posts/time window] the rate with which user $n$ generates new posts, and  we make the assumption that content posted instantaneously appears on the Newsfeeds of his followers and can be further propagated through the social network - depending on the platform. For all users $n \in \mathcal{N}$ we suppose that they keep a constant post rate $\lambda^{(n)}$ during the time-window.

In this work we consider fixed post rates within the observed time-window to limit the analysis in static scenarios and convex optimisation problems. However, as an extension we can generalise our research to cases where the posting rates vary over time. Such analysis can be possible either by discretizing time and assuming the rate constant but possibly different in each slot, or by working with counting processes and continuous control theory. These are interesting extensions for future work, nevertheless, in the real world, we believe that their usefulness will be limited. The reason is that in practice the advertiser needs to negotiate with each influencer separately and make a sort of arrangement over a relatively long period of time (say some weeks or months of posting) before the investment results can be evaluated and the advertising budget re-distributed. During this period (time-window) the advertiser needs to consider the posting rate of influencers constant on average, and cannot break the agreement with the influencers at will. 

At each point in time, a user sees in his (her) Newsfeed posts originated by other users who may or may not be their direct leaders, depending on the type of platform. The number of these viewed posts are called \textit{impressions} and the \textit{impression ratio} is the ratio of the impressions originated by some given user over all viewed impressions in a given snapshot. The average ratio over several snapshots is called the \textit{average impression ratio} in the time window.

Let us denote by $p^{(j)}_n$ the average impression ratio of posts that originate from user $n$ in the Newsfeed of user $j$. This quantity $p^{(j)}_n$ is assumed known for the rest of the article and can be measured or estimated in two ways:
\textit{Empirically}, by taking multiple Newsfeed snapshots in the time window and calculating the average of the ratio of impressions between pairs of users over those time points.
Alternatively, \textit{through Markovian analysis}. Namely, if we assume complete knowledge of the social graph and user posting activity, the values $p^{(j)}_n$ can be derived using the Markovian diffusion model introduced in \cite{a15}.

Naturally, the average impression ratios satisfy for each Newsfeed:
\begin{equation}
\sum_{n \in \mathcal{N}}  p^{(j)}_n=1, \ \forall j \in \mathcal{N}.
\end{equation}

Our model does not require explicit knowledge of the list of followers and leaders of each user, nor a diffusion process as in the approach by Kempe et al. \cite{a1}. However, it does require knowledge over the average impression ratio, that contains all this information resulting from diffusion. 
Furthermore, we are interested in studying the relative impact between pairs of users and not the absolute impact, since the Walls and Newsfeeds can vary in size between users.

Note here that in Instagram and other OSPs, due to the lack of a re-posting option the propagation of information is only given to the immediate followers of a user, thus hindering post-propagation. These networks are simpler to describe; the user sets form a bipartite graph (leaders/followers).

\subsection{The budgeted portfolio optimization problem}

In the budgeted portfolio optimization problem a given fixed advertiser $i \in \mathcal{N}$ with a certain monetary budget $B$ [EUR/time window] at his (her) disposal orchestrates an advertising campaign in a unit of time (equal to the time window) by investing on other users to create posts in his (her) favour. The aim is to maximize some impact metric, e.g. the number of impressions, the sales, or the audience reach.

We suppose that for each user $n\not =i$ there is a known associated cost per post $c_n$ [EUR/post] so that the user $n$ will be willing to create posts in favor of the advertiser $i$.

In order to formulate this optimization problem, we need to quantify the participation of each user $n$ in the campaign of the advertiser $i$. Hence, we define for each user $n\not =i$, the \textit{unknown participation ratio} $a_n \in [0,1]$ in the campaign as the unknown proportion of user $n$'s generated posts acquired by the advertiser $i$ in the unit of time. We fix $a_i=1$ meaning that the advertiser always posts to promote its own product. Then, $a_n \lambda_n$ [posts/time window] represents the number of posts that the user $n$ creates in favor of the advertiser $i$. Note that the unknown participation ratio $a_n$ is dependent on the advertiser $i$, therefore the notation should be $a^{i}_n.$ However, we have assumed that the advertiser is fixed from the beginning of the campaign, so for the sake of clarity and simplicity in the notation, we omit the superscript $i$ and choose to represent $a^{i}_n$ as $a_n$ throughout the manuscript.

Similarly, we define by $p^{(j)}_n(a_n)$ the \textit{campaign-related impression ratio} as the average value of the impression ratio in the Newsfeed of user $j$ originating from user $n$ and related to the campaign of the advertiser $i$. The campaign-related impression ratio can be similarly estimated and measured as above. Note that by construction $p^{(j)}_n(a_n) \leq p^{(j)}_n(1), \forall n,j \in \mathcal{N}$ and by definition $p^{(j)}_n=p^{(j)}_n(1), \forall n,j \in \mathcal{N}$, so we have:
\begin{equation}
p^{(j)}_n(a_n) \leq p^{(j)}_n \ \ \forall n,j \in \mathcal{N}.
\end{equation}

Observe that the impression ratio and the campaign-related impression ratio quantify the heterogeneity of users in the propagation of posts or ads in network.

The empirical probability that an impression reaching user $j$ is campaign-related, is called the \textit{potential of user $j$}:
\begin{eqnarray}
\label{omega}
\omega^{(j)}(\mathbf{a})=\sum_{n \in \mathcal{N}\setminus \{j \} } p^{(j)}_n(a_n) & \leq & 1.
\end{eqnarray}

In the above $\mathbf{a}=(a_1,...,a_{i-1},a_{i+1},...,a_N)^{T}$ is the participation vector of all the users into the advertising campaign of user $i$, excluding user $i$, whose $a_i=1$.

We introduce a utility function $U_j$ for each user $j$ that maps the potential of user $j$, $\omega^{(j)}$, to the campaign objective of the advertiser. Different expressions for $U_j$ model different performance metrics. 

The budget invested to user $n\not =i$ by the advertiser is $ B_n(a_n) = c_n a_n \lambda ^ {(n)}$ [EUR/time window]. Given that the total budget of the advertiser $i$ is $B$ [EUR/time window], the constraints in our Budgeted Portfolio Optimization (BPO) problem will be naturally a budget restriction $\sum_{n \not =i} B_n(a_n) \leq B$ and the continuous unknown variables $a_n \in [0,1]$. Altogether, 
\begin{align*}
\label{[BPO]}
\textrm{max}_{\{a_n \}_{n \not=i}} \quad & U(\mathbf{a}):=\sum_{j \in \mathcal{N}\setminus \{i \} } U_j(\omega^{(j)}(\mathbf{a})),\\
\textrm{ s.t.} \quad & \sum_{n \in \mathcal{N}\setminus \{i \} }  c_n a_n \lambda ^ {(n)} \leq B,\tag*{[BPO]}\\
& a_i=1, \ 0 \leq a_n \leq 1,  \forall n \in \mathcal{N}.
\end{align*}

$U(\mathbf{a})$ is the total utility of the advertiser's campaign.

\subsection{Variations and extensions}
The above formulation allows us to introduce further extensions of our model:

\begin{enumerate}
    \item  We can consider that users want to sell no more than a certain ratio of their posts $a_n \leq r_n  \leq 1, \forall n$.
    \item Another variation is by introducing a set of posting categories to every user $\varsigma_n$. Then an influencer-follower user pair $(n,j)$ is activated, only when the two users share some common interests. In this case the potential of follower $j$ is expressed as: 
    \begin{equation*}
        \omega^{(j)}(\mathbf{a})=\sum_{n \in \mathcal{N}\setminus \{j \} } p^{(j)}_n (a_n) I_{\varsigma_n \cap \varsigma_j \not = \varnothing },
    \end{equation*}
    with $\varsigma_n \subset \{ 1,...,\text{Number \ of \ categories} \}$ the hobbies or interests of user $n$ and similarly for $\varsigma_j$ about user $j$.
\end{enumerate}

\subsection{Assumption on ad propagation and impact metrics}
An assumption for the rest of the article is that we consider a linear propagation for the posts related to the campaign and seen on the Newsfeeds, namely:
\begin{equation}
\label{eq4}
    p^{(j)}_n(a_n)=a_n p^{(j)}_n.
\end{equation}

This is reasonable because if the user $j$ is an immediate follower of influencer $n$, and all posts from the influencer appear on his Newsfeed, then a percentage $a_n$ will be related to the campaign. This is actually the case for platforms without sharing, like Instagram, where a user views every post published from each followee, and the viewed ads will just be the percentage of the posts that the followee creates for the advertiser. In the case of other platforms, impressions can result from diffusion, arriving through sharing of content from intermediate users. Then, the above linear expression implies that a post from $j$ is shared randomly, independently of its content, which of course is not true. We will use however the linear assumption as a reasonable approximation to the campaign diffusion process for any platform, because we lack of any prior information related to how users might react to the campaign’s posts. So, for the rest of the article, the potential of the user $j$ is expressed as: 
\begin{equation}
    \omega^{(j)}(\mathbf{a})=\sum_{n \in \mathcal{N}\setminus \{j \} } a_n p^{(j)}_n.
\end{equation}

The utility function $U_j$ in~\ref{[BPO]} of the user $j$, represents from a modeling point of view the following:

\begin{itemize}
    \item \textit{Impressions/Engagements}: In this case, the objective function for each user is a linear function. This translates as follows: an increase in the impression potential (5) of a user $j$ results in a proportional increase in their utility.
    \item \textit{Conversion/Sales}: 
    In this work we suppose that a concave utility function models diminishing returns over the potential of each user $j$. As the amount of one participation ratio increases, then after some point the marginal conversion/sales (extra output gained by adding an extra unit) decreases. In fact, this comes from microeconomic theory, where utility functions are usually assumed to be concave or quasi-concave over some or all of their domains to incorporate the property of diminishing returns \cite{MSP}. Note that it is possible that the advertising-response function in the marketing context has a shape that is first convex and then concave, i.e. it is S-shaped \cite{ARC}. Such S-shaped curves can also be treated by our algorithm as will be shown in section~\ref{sec:III}.
    \item \textit{Reach}: We model this case by the $\alpha$-fair concave function, when $\alpha\rightarrow \infty$ \cite{a34}, \cite{a35}. This gives a $\max-\min$ solution. Alternatively, we can provide user-specific thresholds $\epsilon_j$ for each user $j$. Then, if the user $j$ sees more than the threshold $\epsilon_j$ campaign-related impression ratio $\omega^{(j)}(\mathbf{a})$, then the user $j$ is consider to be reached by the campaign:
    \begin{equation}
    U_j(\omega^{(j)}(\mathbf{a}))=I_{\omega^{(j)}(\mathbf{a})>\epsilon_j}.
\end{equation}
\end{itemize}

Note that the $\alpha$-fairness family \cite{a34}, \cite{a35} is a general class of concave utility functions with diminishing returns that can capture various fairness criteria such as proportional fairness for $\alpha\rightarrow 1$ (Sales/ROI) and max-min fairness
for $\alpha\rightarrow\infty$ (Reach), and many others with a suitable choice of
$\alpha \in (0,\infty)$. The general expression is 
\begin{eqnarray}
U_j(\omega(j)) = \frac{(1 + \omega^{(j)})^{1-\alpha}}{1-\alpha}.
\end{eqnarray}

Hence, under the assumption of linear propagation (\ref{eq4}) and activity constraints $\{r_n \}_{n \not=i}$ we have the formulation of the general budgeted portfolio optimization problem (BPO-G) in an OSP for various impact objectives:
\begin{align*}
\label{[BPO-G]}
\textrm{max}_{\{a_n \}_{n \not=i}} \quad & U(\mathbf{a}):=\sum_{j \in \mathcal{N}\setminus \{i \} } U_j\left(\sum_{n \in \mathcal{N}\setminus \{j \} } a_n p^{(j)}_n\right), \\
\textrm{ s.t.} \quad & \sum_{n \in \mathcal{N}\setminus \{i \} }  c_n a_n \lambda ^ {(n)} \leq B, \tag*{[BPO-G]}\\
& a_i=r_i, \ 0 \leq a_n \leq r_n,  \forall n \in \mathcal{N}. 
\end{align*}

Observe that the feasibility set for~\ref{[BPO-G]} is:

\begin{align}
\label{eqfeas}
\mathcal{M}=\{ \mathbf{a}\in \mathbb{R}^{N-1} | \sum_{n \in \mathcal{N}\setminus \{i \} }  c_n a_n \lambda ^ {(n)} \leq B, \nonumber\\
\forall n \in \mathcal{N}\setminus \{i \}, \ 0 \leq a_n \leq r_n\},
\end{align}

Therefore, $\mathcal{M}$ is a convex and compact set. 
Moreover, let us consider the set of points $\mathcal{S}$ in $\mathcal{M}$ satisfying the condition $\sum_{n \in \mathcal{N}\setminus \{i \} }  c_n a_n \lambda ^ {(n)}=B,$ and $E(\mathcal{S})$ the set of extreme points of $\mathcal{S}.$ In \cite{I0} we have characterised the extreme points of $\mathcal{S}$, and made a relation with the global optimizer of the problem~\ref{[BPO-G]}. This is summarised as follows:

\begin{proposition}
\label{propp} (From \cite{I0}).  
(i) Let $(a_n)_{n \not=i} \in E(\mathcal{S})$ be an extreme point of $\mathcal{S}$ in [BPO-G], then it satisfies the next property: there exists at most one $j \in \mathcal{N}\setminus \{i \}$ such that $a_j \in (0,r_j)$ and $\forall l \in \mathcal{N}\setminus \{i,j \}, \  a_l \in \{0,r_l \}$. Conversely, any point $\{a_n \}_{n \not=i} \in \mathcal{S}$ that satisfies this property is an extreme point. (ii) Furthermore, a global maximizer in [BPO-G] can be written as a convex combination of points satisfying this attribute.
\end{proposition}


Notation and parameters are summarized in Table \ref{table:2}.

\begin{table}[!ht]
\centering
\resizebox{\columnwidth}{!}{
\begin{tabular}{||c c||} 
 \hline
 \multicolumn{2}{||c||}{Budgeted Portfolio Optimization Problem}  \\[0.5ex] 
 \hline\hline
 User set & $\mathcal{N}$ \\ 
 Advertiser & $i \in \mathcal{N}$\\ 
 Average impression ratios & $\{p^{(j)}_n \}_{n, j \in \mathcal{N}}$  \\
 Cost per post & $\{c_n \}_{n \in \mathcal{N}}$ \\
 Budget over a time period & $B$ \\
 Activity restriction (optional) & $\{r_{n} \}_{n \in \mathcal{N}}$ \\ 
 User activity & $\{\lambda^{(n)} \}_{n \in \mathcal{N}}$ \\ [1ex] 
 \hline\hline
  \multicolumn{2}{||c||}{Objective:}  \\
 \hline
 Campaign objective & $U(\mathbf{a}):=\sum_{j \in \mathcal{N}\setminus \{i \} } U_j(\sum_{n \in \mathcal{N}\setminus \{j \} } [a_n p^{(j)}_n])$\\  [1ex] 
 \hline\hline
 \multicolumn{2}{||c||}{Return:}  \\
 \hline
 Participation ratio per user & $\mathbf{a}:=\{a_n \}_{n \not=i}$\\
  \hline
\end{tabular}}
\caption{Elements of our problem and notation}
\label{table:2}
\end{table}

\section{Solution to the advertiser's campaign}
\label{sec:III}

In this section, we use the above properties of the feasible set to propose one low-complexity fast algorithm to solve the optimization problem~\ref{[BPO-G]}. This algorithm has a space complexity in memory of order $\mathcal{O}(max(N-1,D))$ and an average-computational complexity in time of order $\mathcal{O}(max((N-1) \log(N-1),D))$ per iteration, with an $\epsilon$-approximate solution to the problem in the Primal after $\mathcal{O}(1/\epsilon)$ steps. Besides, we discuss as a corollary, the case of linear utility functions as analysed in \cite{I0}, and consequently we propose a general heuristic for the solution of~\ref{[BPO-G]}. 


\subsection{Frank-Wolfe algorithm}
The standard in convex optimization would be to translate a problem into a conic program and solve it using a primal-dual interior point method. However, in general, the iteration costs of interior point methods grow super-linearly with the dimension of the problem \cite{III0,III01,III1}. As the dimension $N$ of our optimization problems can be very large, interior point methods eventually become impractical and we require alternative methods. Particularly attractive to this aim are the first order methods \cite{IIIA1}. One of the earliest such methods was introduced by Frank-Wolfe in 1956 \cite{III12} to solve the optimization problem:
\begin{equation}
\label{eq1}
\begin{aligned}
\textrm{max}_{\mathbf{a} \in \mathcal{M}} U(\mathbf{a}),
\end{aligned}
\end{equation}
with $U:\mathcal{M} \subset \mathbb{R}^{N-1} \xrightarrow{} \mathbb{R}$ a continuously differentiable function that is concave and with $\mathcal{M}$ convex and compact.

The Frank-Wolfe (FW) algorithm has recently re-emerged due to its applications in numerous fields \cite{IIIFA1,IIIFA2,IIIFA3}. It constructs a sequence of estimates $\mathbf{a}^{(1)},\mathbf{a}^{(2)},...$ that converges towards the solution of the optimization problem~(\ref{eq1}) given some initial guess $\mathbf{a}^{(0)}$. The FW mechanism relies on a routine that iteratively formulates and solves linear problems over the domain $\mathcal{M}$. This routine is commonly referred to as a linear maximization oracle \cite{III22}. Hence, in each iteration, the Frank–Wolfe algorithm considers a linear approximation of the objective function, and moves towards a maximizer in $\mathcal{M}$ in the direction dictated by the oracle $\mathbf{d}_t=\mathbf{s}^{(t)}-\mathbf{a}^{(t)}$.

The Frank-Wolfe gap after $t$ iterations is defined by $\mathbf{g}_t= \langle \nabla U(\mathbf{a}^{(t)}), \mathbf{d}_t \rangle$. Observe that when $U$ is a concave function, by definition, the Primal gap after $t$ iterations satisfies $U(\mathbf{a}^{*})-U(\mathbf{a}^{(t)}) \leq \mathbf{g}_t, \ \forall t \in \mathbb{N},$ with $\mathbf{a}^{*} \in \mathcal{M}$ a stationary point of $U$ (i.e. $\langle \nabla U(\mathbf{a}^{*}), \mathbf{a}^{*}-\mathbf{a} \rangle \geq 0, \ \forall \mathbf{a} \in \mathcal{M}$). So, the Frank-Wolfe gap provides us with a tool to find $\epsilon$-approximations to the optimum of the problem in~(\ref{eq1}).

In the case that $U$ is concave with finite curvature $C$ in $\mathcal{M}$, then, as we will formally show later, the Frank-Wolfe algorithm converges with rate $\mathcal{O}(1/t)$ of both the Primal and the Frank-Wolfe gap to the optimum, where $t$ are the iterations. As described in \cite{III2,III22}, this rate of convergence of the Frank-Wolfe algorithm 
is also achieved for search step size either $\gamma_t = \argmin_{\gamma \in [0, 1]} U(\mathbf{a}^{(t)} + \gamma \boldsymbol{d}_t)$ or $\gamma_t = \vphantom{\sum_i}\min\Big\{\frac{\mathbf{g}_t}{C}, 1 \Big\}.$


The result can be extended for $U$ continuously differentiable with finite curvature in $\mathcal{M}$ but non-concave. Then the Frank-Wolfe algorithm converges with rate $\mathcal{O}(1/t)$ of the Primal and with rate $\mathcal{O}(1/\sqrt{t})$ of the minimal Frank-Wolfe gap after $t$ iterations as shown in \cite{III22,III31}. By extending the gradient concept to sub-gradients of $U$ in the case that $U$ is concave but non-smooth, we have analogous results.

\subsection{Our adaptation}

In order to apply the Frank-Wolfe algorithm to the optimization problem~\ref{[BPO-G]}, it is needed first to compute the derivative of the total utility defined by the problem~\ref{[BPO-G]}, which, by definition is $\nabla_k U(\mathbf{a}^{(t)})=\sum_{j \in \mathcal{N}\setminus \{i,k \}} U'_j(\omega^{(t)}_j) p^{(j)}_k.$ Subsequently, note that it is necessary for each step to find $\mathbf{s}^{(t)}$ through the conditional gradient problem $\mathbf{s}^{(t)} \in \argmax_{\mathbf{s}^{(t)} \in \mathcal{M}} \langle \nabla U(\mathbf{a}^{(t)}), \mathbf{s}^{(t)} \rangle$, and hence the next direction $\mathbf{d}_t=\mathbf{s}^{(t)}-\mathbf{a}^{(t)}$. 

At this point, we exploit the special structure of our optimization problem~\ref{[BPO-G]} to formulate the conditional gradient problem as a linear program with a low-complexity fast solution for each step. The solution of the linear program that arises is found by sorting as the next Lemma~\ref{lema2} proved in the Appendix~\ref{appendix:a} states. We can apply  Lemma~\ref{lema2} to solve the conditional gradient problem by setting $K_{j}=\nabla_j U(\mathbf{a}^{(t)})$ for each step $t$. This approach is described in Algorithm~\ref{algo3}.

Notice that, compared to the standard Frank-Wolfe method, our algorithm actually has an important modification that does not affect the convergence rate, but it does however improve the complexity (time and memory) to calculate each iteration of the algorithm. Specifically, this is presented in Lemma~\ref{lema2}. Based on the specific structure of the constraint set, we found that the Linear Program involved in each step of the process is not necessary to be solved using the standard simplex method; but a simple ordering of the quantities $\frac{\nabla_k U(\mathbf{a}^{(t)})}{c_k\lambda^{(k)}}$ and a selection of participation ratios according to Algorithm~\ref{algo3} finds the exact same solution much faster. This corresponds to a reduction in complexity, from $O(2^N)$ in worst case or polynomial on average in the simplex algorithm case, to $O(Nlog(N))$ for our Algorithm~\ref{algo3}. This corresponds to a significant improvement in complexity, which is crucial for practical purposes especially for real world problems with very large number of users.



\begin{lemma}
\label{lema2}
Let us consider the vector $\rho \in \mathbb{R}_{\geq 0}^{N-1}$ defined as $\rho=(c_1 \lambda^{(1)},...,c_{i-1} \lambda^{(i-1)},c_{i+1} \lambda^{(i+1)},...,c_N \lambda^{(N)}),$ $\mathbf{s}=(s_1,...,s_{i-1},s_{i+1},...,s_N) \in \mathcal{M}$, $K=(K_1,...,K_{i-1},K_{i+1},...,K_N) \in \mathbb{R}^{N-1}$, the budget $B \geq 0$, and the user set $\mathcal{Z}=\{l \in \mathcal{N}\setminus \{i \} | \rho_l \not =0\}$. With these let us formulate the following linear program: 
\begin{align*}
\label{maxlin}
\textrm{max}_{\mathbf{s}} \quad & \sum_{j \in \mathcal{N}\setminus \{i \} } K_j  s_j, \\
\textrm{ s.t.} \quad & \sum_{n \in \mathcal{N}\setminus \{i \} }  \rho_n s_n \leq B, \tag*{\textit{[LP]}}\\
& 0 \leq s_n \leq r_n,  \forall n \in \mathcal{N}\setminus \{i \}. 
\end{align*}

Suppose the users $\{i_{k} \}_{k=1,...,|\mathcal{Z}|}$ are indexed in decreasing order of their quantities $\frac{K_{i_{k}}}{\rho_{i_k}}$. Furthermore, $\tau$ is defined as the maximum index from the ordered users such that $\sum^{\tau}_{m=1}  \rho_{i_{m}} r_{i_{m}} \leq B.$ Then a solution $\mathbf{s}^{*}$ to ~\ref{maxlin} is given for all $j \in \mathcal{N}\setminus \{i \}$ as:
\[
  s^{*}_j =
  \begin{cases}
                                r_j I_{K_j > 0} & \text{if $j \notin \mathcal{Z}$ and $j \not =i$}, \\
                                   r_{j} & \text{if $j=i_k \in \mathcal{Z}$ and $k \leq \tau$}, \\
                                   \frac{B-\sum^{\tau}_{\ell=1} \rho_{i_{\ell}} r_{i_{\ell}}}{\rho_{i_{\tau+1}}}
                                   & \text{if $j=i_k \in \mathcal{Z}$ and $k=\tau+1$},\\
                                   0 & \text{if $j=i_k \in \mathcal{Z}$ and $k \geq \tau+2$}.
  \end{cases}
\]
\end{lemma}

\begin{proof}
The proof can be found in the Appendix~\ref{A-1}.
\end{proof}

\RestyleAlgo{boxruled}
\begin{algorithm}[tbh]
 
 \textbf{Initialization}. Let $\mathbf{a}^{(t)}$ and $\nabla U(\mathbf{a}^{(t)})$ from the current iteration $t$ in the outer loop of Algorithm~\ref{algo2}.
        
        Compute users $\mathcal{Z}=\{n \in \mathcal{N}\setminus \{i \}  | c_n \lambda ^ {(n)} \not =0 \}$\;

        $\{i_{k} \}_{k \in \{ 1,...,|\mathcal{Z}| \}} = \textbf{Index Sort} \{\frac{\nabla_k U(\mathbf{a}^{(t)})}{c_k \lambda ^ {(k)}}  \}_{k \in \mathcal{Z}}$\;

        $\tilde{B}=B$\;
        
        \For{$n \in \mathcal{N}\setminus \mathcal{Z}$}{$\mathbf{s}_n^{(t)}=r_n I_{\nabla_n U(\mathbf{a}^{(t)})>0}$}

        \For{$z=1,...,Z$}{
        
        \eIf{\textit{ $c_{i_z} r_{i_z} \lambda ^ {(i_z)} \leq \tilde{B}$}}{
      $\mathbf{s}_{i_z}^{(t)}=r_{i_z}$
      \;
    }{
      $\mathbf{s}_{i_z}^{(t)}=\frac{\tilde{B}}{c_{i_z} \lambda ^ {(i_z)}}$
      \;
    }
    
    $\tilde{B} \gets \tilde{B}-\mathbf{s}_{i_z}^{(t)} c_{i_{z}} \lambda ^ {(i_{z})}$
    
    }

 \KwResult{$\mathbf{s}^{(t)}$}
 \caption{Solution to the Linear sub-problem.}
 \label{algo3}
\end{algorithm}

We proceed to adapt the Frank-Wolfe algorithm using Algorithm~\ref{algo3} for the special case of~\ref{[BPO-G]} to propose a low-complexity fast algorithm described in Algorithm~\ref{algo2}. 

\begin{theorem}
\label{teo1}
Consider the optimization problem~\ref{[BPO-G]} with $U$ a continuously differentiable function that is concave with finite curvature $C$ and let $D=\{(n,j) \in \mathcal{N} \times \mathcal{N} | p^{(j)}_{n} \not = 0\}$ and $N=|\mathcal{N}|$. Then Algorithm~\ref{algo2} converges with rate $\mathcal{O}(1/t)$ of both the Primal and the Frank-Wolfe gap after $t$ iterations to the optimal solution, for any variant of step size $\gamma_t=\frac{2}{t+2},$ or  $\vphantom{\sum_i}\min\Big\{\frac{\mathbf{g}_t}{C}, 1 \Big\},$ namely: 
\begin{equation}
    \label{ceq25}
    U(\mathbf{a}^{*})-U(\mathbf{a}^{(t)})\leq \frac{2C}{t+2}, \ \forall t \geq 0,
\end{equation}
\begin{equation}
    \label{ceq26}
    \mathbf{g}_t \leq \frac{2\beta C }{t+2}, \ \forall t \geq 0, \ T \geq 2,
\end{equation}
where $\mathbf{a}^{*}$ is an optimal solution to the optimization problem~\ref{[BPO-G]} and $\beta=\frac{27}{8}$.

Moreover, Algorithm~\ref{algo2} has a computational complexity in memory of order $\mathcal{O}(max(N-1,D)).$ Additionally, Algorithm~\ref{algo2} has an average- and worst-computational complexity per step in time of order $\mathcal{O}(max((N-1) log(N-1),D)),$ and a best-computational complexity per step in time of order $\mathcal{O}(max(N-1,D)).$
\end{theorem}

\begin{proof}
The proof can be found in the Appendix~\ref{A-2}.
\end{proof}

\RestyleAlgo{boxruled}
\begin{algorithm}[tbh]
 
 \textbf{Initialization}. Let $\mathbf{a}^{(0)} \in \mathcal{M}$, maximum iterations $T$ and tolerance $\epsilon>0$.
 
 \For{\textcolor{black}{$t=0,...,T$}}{
 
 Update $\omega^{(t)}$ defined as $\omega^{(t)}_j=\sum_{n \in \mathcal{N}\setminus \{j \} } \mathbf{a}^{(t)}_n p^{(j)}_n$\;
 
    Update $\nabla U(\mathbf{a}^{(t)})$ defined as $\nabla_k U(\mathbf{a}^{(t)})=\sum_{j \in \mathcal{N}\setminus \{i,k \}} U'_j(\omega^{(t)}_j) p^{(j)}_k$\;
 
  Compute $\mathbf{s}^{(t)}$ by Algorithm~\ref{algo3}\;
  
  Let $\mathbf{d}_t=\mathbf{s}^{(t)}-\mathbf{a}^{(t)}$\;
  
  Compute $\mathbf{g}_t= \langle \nabla U(\mathbf{a}^{(t)}), \mathbf{d}_t \rangle$\;

  \eIf{$\mathbf{g}_t<\epsilon$}{\textbf{return} $\mathbf{a}^{(t)}$}{
  Set the step size $\gamma_t$\;
  Update $\mathbf{a}^{(t+1)}=\mathbf{a}^{(t)}+\gamma_t \mathbf{d}_t$\;
  }

  Set $t \xleftarrow[]{} t+1$\;
 }
 \KwResult{$\mathbf{a}^{(t)}$}
 \caption{Frank-Wolfe applied to~\ref{[BPO-G]}.}
 \label{algo2}
\end{algorithm}

The computational complexity in time and memory depends on the algorithm selected to sort the users of Algorithm~\ref{algo3}. For this reason \texttt{timsort} is used due to its advantage in complexity and stability over other algorithms such as \texttt{mergesort} \cite{III6}. Hence, Algorithm~\ref{algo2} and Algorithm~\ref{algo3} are fast in sparse environments where $D \ll N^{2}.$ It is sufficient to calculate $\mathcal{Z}$ only once at the beginning of Algorithm~\ref{algo2} and not at each run of Algorithm~\ref{algo3}.




Theorem~\ref{teo1} can be generalised to environments where a linear propagation model defined by equation~(\ref{eq4}) does not hold (as generally described in~\ref{[BPO]}), but it has a known expression. In this case, it will be sufficient in the Algorithm~\ref{algo2} to update $\omega^{(t)}_j$ and $\nabla_k U(\mathbf{a}^{(t)})$ as $\sum_{n \in \mathcal{N}\setminus \{j \} } U_j[p^{(j)}_n(\mathbf{a}^{(t)}_n)]$ and $\sum_{j \in \mathcal{N}\setminus \{i,k \}} U'_j(\omega^{(t)}_j) [(p^{(j)}_k)'(\mathbf{a}^{(t)}_n)]$ respectively. Furthermore, in the case that we have S-shaped advertising-response curves, Theorem~\ref{teo1} can be extended with convergence rate of the primal $\mathcal{O}(1/t)$ and with rate $\mathcal{O}(1/\sqrt{t})$ of the minimal Frank-Wolfe gap after $t$ iterations as shown in \cite{III22,III31}.

Note that Algorithm~\ref{algo2} and Algorithm~\ref{algo3} match with our earlier version of this paper by taking a step size of $1/(t+2)$ instead of $2/(t+2)$.


\subsection{Rule-of-thumb}
\label{rule}
Note here that the Frank-Wolfe solution of~\ref{[BPO-G]} for linear utilities can be derived in just one step of Algorithm~\ref{algo2}. This observation matches with the conference version of this paper for the linear case, which is again found in one step, see Reference \cite{I0}. From Algorithm~\ref{algo3} the users are ordered in decreasing order of the quantity $\nabla U(\mathbf{a})/c_k \lambda ^ {(k)}$, and in the special case of linear utilities 
we obtain a simple and practical rule-of-thumb: First, order the users in decreasing order of the quantity $\{\frac{\phi_k}{c_k \lambda ^ {(k)}} \}_{k \in \mathcal{N}\setminus \{i \}}$, where $\phi_k = \sum_{j\in\mathcal{N}\setminus \{k,i\}}p_k^{(j)}$. Then, select influencers from top to bottom, investing in their full activity, until the budget is exhausted. This way, all the selected influencers, with one possible exception (the last in the selection) will have $a_j=1$ (or $r_j$ if limited) participation ratio. The rule is exact for linear utilities and approximative in other cases: this rule of influencer choice balances high influence with low cost per post and total activity.

\section{Multi-instances
} 
\label{sec:IV}



We proceed to generalise the problem formulation in the case of $L\geq 1$ generic OSPs forming the set $\mathcal{L}$, with $Q\geq 1$ different content-types posted (e.g. text, image or video), described as the set $\mathcal{Q}$. We can assume without loss of generality that all users are present in all platforms and can post any content; this will further be explained later. 

The rate with which user $n \in \mathcal{N}$ generates new posts of content $q$ on the platform $l$ is denoted by $\lambda_{l,q,n}$ [posts/time window]; the rate is assumed fixed within the window of observation. 
For each platform $l$, we denote by $p^{(j)}_{l,q,n}$ the average impression ratio of type $q\in\mathcal{Q}$ posts found on the Newsfeed of user $j \in \mathcal{N}$, and which originate from user $n \in \mathcal{N}$ . These quantities $p^{(j)}_{l,q,n}$, viewed by user $j$ are assumed measured or estimated, as we have already discussed in section \ref{sec:II}. 

Then, if some user $n$ is not present in some platform $l$, the posting rate and average impression ratio can be modelled respectively by setting $\lambda_{l,q,n}=0, \ \forall q\in\mathcal{Q}$ and $p^{(j)}_{l,q,n}=p^{(n)}_{l,q,j}= 0, \ \forall j\in\mathcal{N}$. Similarly, if some category $q$ is not present in some platform $l$, we can model the posting rate and average impression ratio respectively as $\lambda_{l,q,n}=0, \ \forall n \in \mathcal{N}$ and $p^{(j)}_{l,q,n}=p^{(n)}_{l,q,j}= 0, \ \forall n,j \in \mathcal{N}.$ Hence, we can assume without loss of generality that each content and user are physically present in each platform.

Naturally, our average impression ratios on the Newsfeed of any user $j\in \mathcal{N}$ and for each platform $l\in\mathcal{L}$ satisfy:
\begin{equation}
\sum_{q\in\mathcal{Q}} \sum_{n \in \mathcal{N}}  p^{(j)}_{l,q,n}=1, \ \forall j \in \mathcal{N},\ \ \forall l\in\mathcal{L}.
\end{equation}

Let us notice that we have a relation with the previous model without content categories and platforms. For this purpose, observe that $\sum_{q\in\mathcal{Q}} \lambda_{l,q,n}$ is the total posting rate of user $n$ in platform $l$. 
Similarly, $\sum_{q\in\mathcal{Q}} p^{(j)}_{l,q,n}$ is the total influence of user $n$ on user $j$ in platform $l$.

\subsection{Multi-platform BPO with multiple content-types}



For each platform $l$ and content $q$, we suppose that each user $n \in \mathcal{N}\setminus \{i \}$ has an associated cost per post $c_{l,q,n}$ [EUR/post] of content $q$. 
It is expected for each user to charge different prices depending on the content-type she (he) produces and the operating platform. In practice, e.g. a video post on Instagram will have a higher price than some text on Twitter. We remark here that posting price varies over platforms due to different performance in ROI.

In order to formulate this optimization problem, we need to quantify the participation of each user $n$ in the campaign of the advertiser $i$. Hence, for each platform $l$ and content $q \in \mathcal{Q}$, we define for each user $n \in \mathcal{N}\setminus \{i \}$, the \textit{continuous participation ratio} $a_{l,q,n} \in [0,1]$ in the campaign as the unknown proportion of user $n$'s generated posts of content $q$ acquired by the advertiser $i$ in the unit of time. In the case that the advertiser is a user on the platform $l$ ($i \in \mathcal{N}$) we fix $a_{l,q,n}=1$ for each content $q \in \mathcal{Q}$ meaning that the advertiser always posts to promote its own product on all platforms. Then, $\lambda_{l,q,n} a_{l,q,n}$ [posts/time window] represents the number of posts of type $q$ that the user $n$ creates in favor of the advertiser on the platform $l$. Furthermore, we can consider that certain users want to sell no more than a certain ratio of their different post contents $a_{l,q,n} \leq r_{l,q,n}  \leq 1, \forall l \in \mathcal{L}, \forall q \in \mathcal{Q}, \forall n \in \mathcal{N}\setminus \{i \}$.

Similarly, we define by $p^{(j)}_{l,q,n}(a_{l,q,n})$ the \textit{campaign-related impression ratio} of content $q$ on the platform $l$ as the average value of the impression ratio of content $q$ in the Newsfeed of user $j$ originating from user $n$ and related to the campaign of the advertiser $i$.  
The empirical probability in the platform $l \in \mathcal{L}$ that an impression related to the campaign reaches user $j \in \mathcal{N}$, regardless of the content, is called the \textit{potential of user} $j$ in the platform $l$:
\begin{equation}
\label{omega2}
\omega_l^{(j)}(\mathbf{a}_{-j};\mathbf{\zeta}_l)= \sum_{q \in \mathcal{Q}} \zeta_{l,q} \sum_{n \in \mathcal{N}\setminus \{j \}} p^{(j)}_{l,q,n}(a_{l,q,n}) \leq 1.
\end{equation}


In the above $\mathbf{a}_{-j}=\{ a_{l,q,n} \}_{l \in \mathcal{L}, q \in \mathcal{Q}, n \in \mathcal{N}\setminus \{j \}}$ is the participation vector of all the users into the advertising campaigns of user $i$. The set of non-negative constants $\zeta_{l}=\{ \zeta_{l,q} \}_{q \in \mathcal{Q}}$ for any $l \in \mathcal{L}$ depends on the relative importance of the posts of content $q$ in the platform $l$, and how these affect the ROI. 
In practice, the set of constants $\{ \zeta_{l,q} \}_{q \in \mathcal{Q}}$ needs to be tuned and if each content has the same relevance in a platform $l$, then we have that $\zeta_{l,q}=\frac{1}{Q}$.



We introduce a utility function $U_{j}$ and the set of non-negative constants $\{ \sigma_{l} \}_{l \in \mathcal{L}}$ that maps the potential of user $j$ in the platform $l$, to the campaign objective $\sigma_l U_{j}(\omega_l^{(j)})$ in the platform $l$ of the advertiser $i$. Different expressions for $U_{j}$ model different performance metrics. Similarly, the set of non-negative constants $\{ \sigma_{l} \}_{l \in \mathcal{L}}$ depends in the relative importance of the platform $l$, it needs to be set by the advertiser to guide the focus of the campaign.


On the platform $l$ the budget invested to user $n \in \mathcal{N}\setminus \{i \}$ for posting content $q$ is $B_{l,q,n}(a_{l,q,n}) = c_{l,q,n} a_{l,q,n} \lambda_{l,q,n}$ [EUR/time window] and the total budget of the advertiser $i$ is $B$ [EUR/time window]. Therefore, the constraints in our budgeted portfolio optimization problem will be naturally a budget restriction $ \sum_{l \in \mathcal{L}} \sum_{q \in \mathcal{Q}} \sum_{n \in \mathcal{N}\setminus \{i \}}  B_{l,q,n}(a_{l,q,n}) \leq B$ and the continuous unknown variables $a_{l,q,n} \in [0,r_{l,q,n}]$. 

Altogether, we can formulate the general budgeted portfolio optimization problem:
\begin{align*}
\label{[M-BPO]}
\textrm{max}_{\{a_{l,q,n}\}_{n \not=i}} \quad & U(\mathbf{a}):=\sum_{l \in \mathcal{L}} \sum_{j \in \mathcal{N}\setminus \{i \}} \sigma_l U_{j}(\omega_l^{(j)}(\mathbf{a}_{-j};\zeta_l)), \tag*{[M-BPO]}\\
\textrm{subject to} \quad & \sum_{l \in \mathcal{L}} \sum_{q \in \mathcal{Q}} \sum_{n \in \mathcal{N}\setminus \{i \}} c_{l,q,n} a_{l,q,n} \lambda_{l,q,n} \leq B,\\
&  0 \leq a_{l,q,n} \leq r_{l,q,n},  \forall l \in \mathcal{L}, \forall q \in \mathcal{Q}, \forall  n \in \mathcal{N},\\
& a_{l,q,i}=1, \forall l \in \mathcal{L}, \forall q \in \mathcal{Q}.
\end{align*}

Above $\mathbf{a}=\{ \mathbf{a}_{-j} \}_{j \not = i}$ is the vector of participation ratios
and $U(\mathbf{a})$ is the total utility of the advertiser's campaign. The new optimization problem~\ref{[M-BPO]} consists of $L Q (N-1)$ variables and the feasibility set is compact and convex.

A linear propagation model is assumed for the posts regardless of the platform and content-type as:
\begin{equation}
\label{eqlpm}
    p^{(j)}_{l,q,n}(a_{l,q,n})=a_{l,q,n} p^{(j)}_{l,q,n}, \forall l \in \mathcal{L}, \forall q \in \mathcal{Q}, \forall  n,j \in \mathcal{N}.
\end{equation}

Using the linear propagation assumption~(\ref{eqlpm}), we can now apply Lemma~\ref{lema2} for the new optimization problem~\ref{[M-BPO]} and obtain an algorithm similar to Algorithm~\ref{algo3} in $L Q (N-1)$ variables. Therefore, a result analogous to Theorem~\ref{teo1} can be derived for $U$ continuously differentiable and concave with finite curvature. The iterative algorithm for the case of multiple platforms and multiple content types is given in the Appendix \ref{a:3}.

Our modeling approach to formulate ~\ref{[M-BPO]} allows for high versatility. For instance, suppose we want instead that~\ref{[M-BPO]} takes into account the fact that a user can be influenced by the ensemble of ads viewed in all platforms, but consequently decide to interact with an ad on a single platform. Then in~\ref{[M-BPO]}, we can instead use a single utility function per user $j \in \mathcal{N}\setminus \{i \}$, having as argument the sum of potentials over all platforms $U_j(\sum_{l \in \mathcal{L}}  \omega_l^{(j)}(\mathbf{a}_{-j};\mathbf{\zeta}_l)),$ and hence the objective function in~\ref{[M-BPO]} changes to $U(\mathbf{a}):=\sum_{j \in \mathcal{N}\setminus \{i \}} U_j(\sum_{l \in \mathcal{L}}  \omega_l^{(j)}(\mathbf{a}_{-j};\mathbf{\zeta}_l)).$ This way, due to the diminishing returns assumption on the shape of $U_j$, multiple ad-publications on various platforms will not have an additive effect on the performance, but all viewed impressions over different platforms will have a group effect in the return. Similarly as above and in the Appendix \ref{a:3}, we can obtain a similar algorithm to Algorithm~\ref{algo3} and an analogous to Theorem~\ref{teo1} can be derived for $U$ continuously differentiable and concave with finite curvature.

\section{Numerical evaluations} 
\label{sec:V}

In this section we evaluate the performance of our models introduced in section~\ref{sec:II} for a single platform and in section~\ref{sec:IV} for multiple platforms. First, we introduce synthetic networks to analyze the impact of graph structure on the campaign. We discuss and compare our Algorithm~\ref{algo2} and the rule of thumb introduced in section~\ref{sec:III} against alternative algorithms from the literature to analyze their comparative advantage for the specific budgeted portfolio optimization problem. Furthermore, we argue how Nano-, Micro-, and Macro-influencers differ in each of the various graph structure across different algorithms, following the description of section~\ref{sec:II}. We numerically investigate the impact of each type of influencers on the different types of networks.


Subsequently, we use a real large Twitter data-log \cite{a54} to evaluate the performance of our Algorithm~\ref{algo2}, for various campaign objectives introduced in section~\ref{sec:III}. Finally, we perform a sensitivity analysis of the multi-platform model introduced in section~\ref{sec:IV}, to analyze the potential ROI-ratio resulting from an optimal budget allocation between $L=2$ networks, one synthetic and the real Twitter data trace \cite{a54}.


\subsection{Numerical simulation}
\label{sec:V-A}

\begin{figure*}[!t]
     \centering
     \begin{subfigure}[b]{0.2\textwidth}
         \centering
         \includegraphics[width=1.1\textwidth]{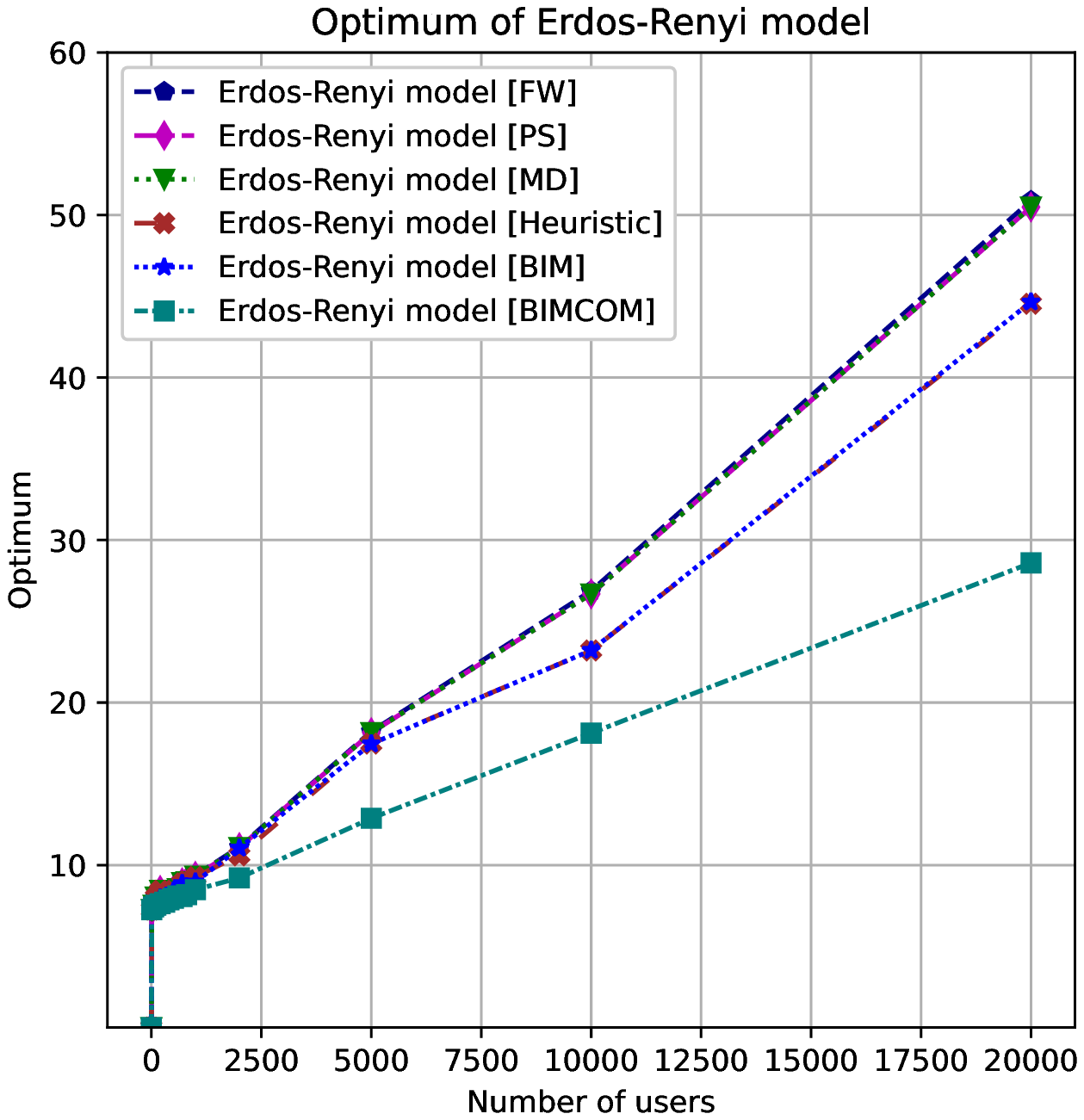}
     \end{subfigure}
     \hfill
     \begin{subfigure}[b]{0.2\textwidth}
         \centering
         \includegraphics[width=1.1\textwidth]{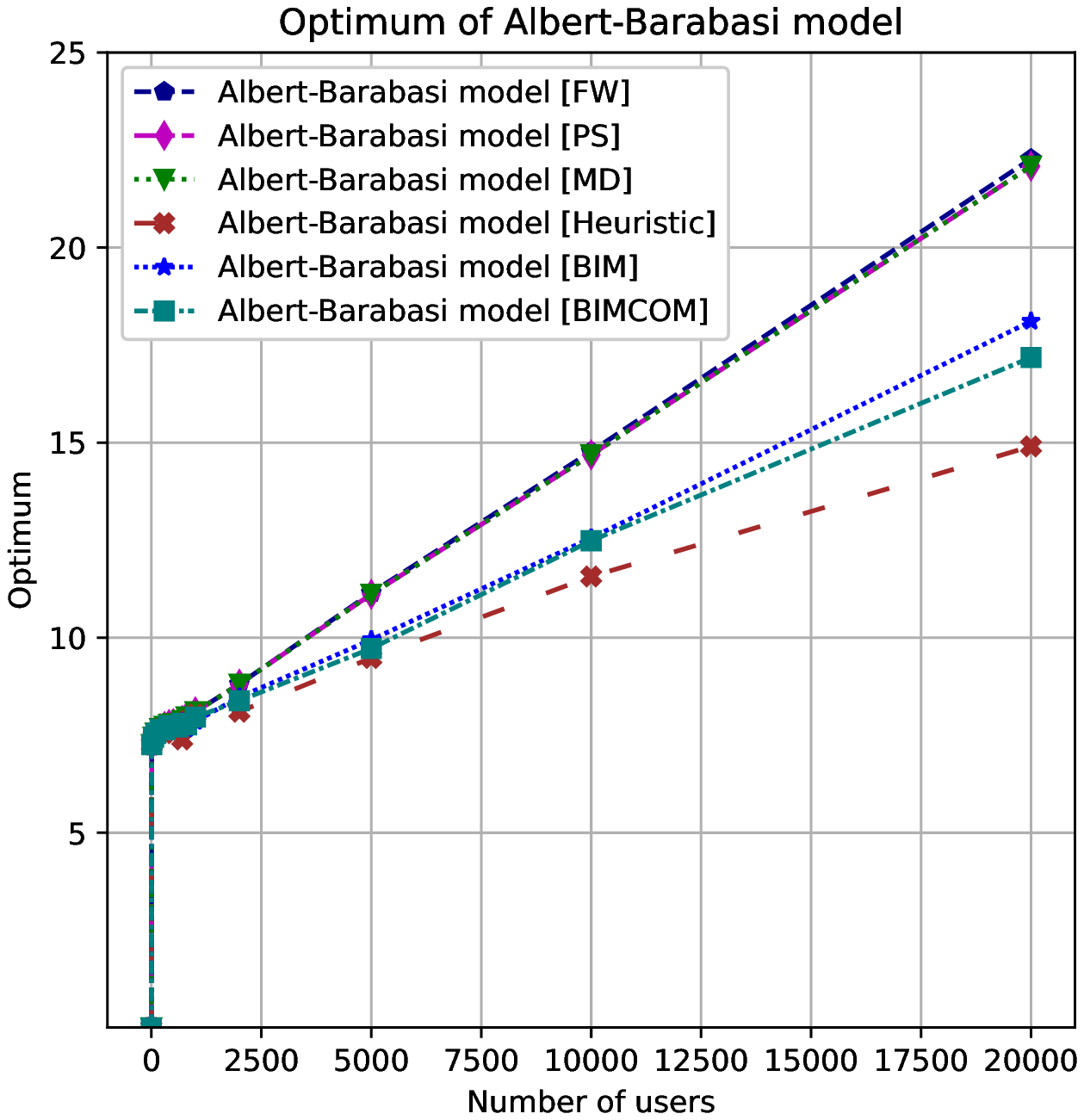}
     \end{subfigure}
     \hfill
     \begin{subfigure}[b]{0.2\textwidth}
         \centering
         \includegraphics[width=1.1\textwidth]{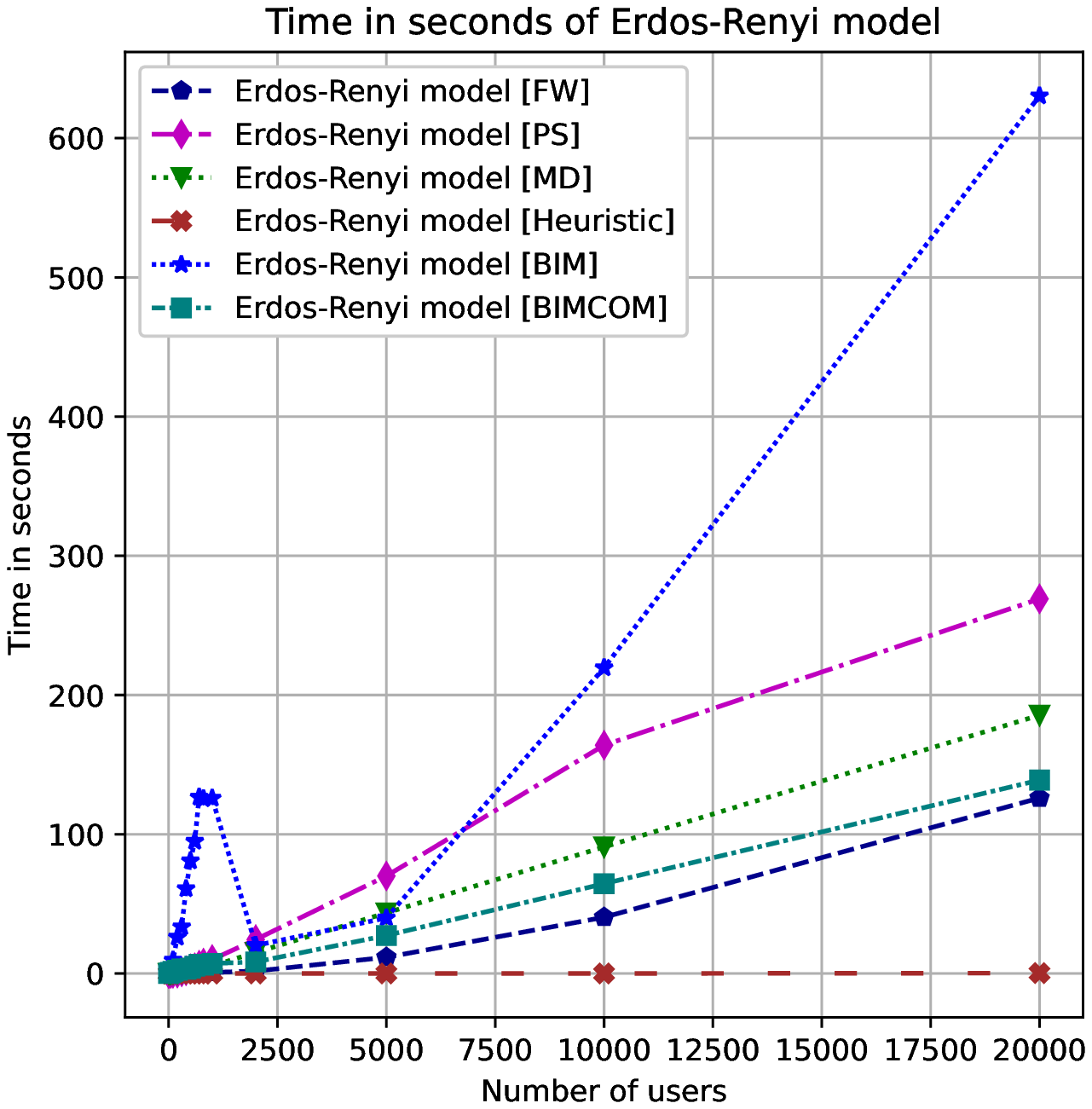}
     \end{subfigure}
     \hfill
     \begin{subfigure}[b]{0.2\textwidth}
         \centering
         \includegraphics[width=1.1\textwidth]{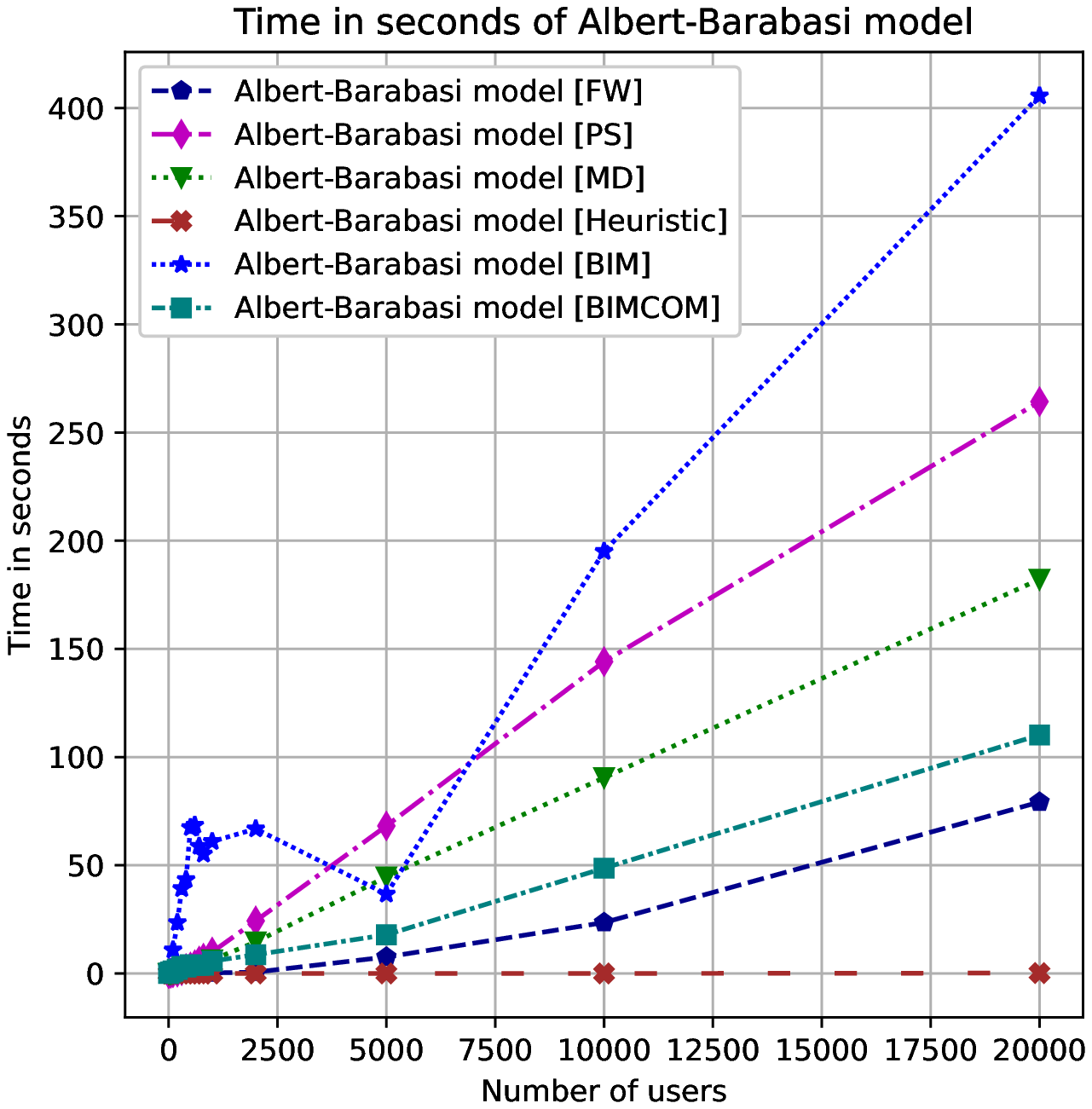}
     \end{subfigure}
        \caption{Synthetic networks and performance metrics for different solution algorithms: (i) Optimum for ER network, (ii) Optimum for AB network (iii) Runtime in ER, (iv) Runtime in AB.}
        \label{fig2}
\end{figure*}

\begin{figure*}[!t]
     \centering
     \begin{subfigure}[b]{0.2\textwidth}
         \centering
         \includegraphics[width=1.1\textwidth]{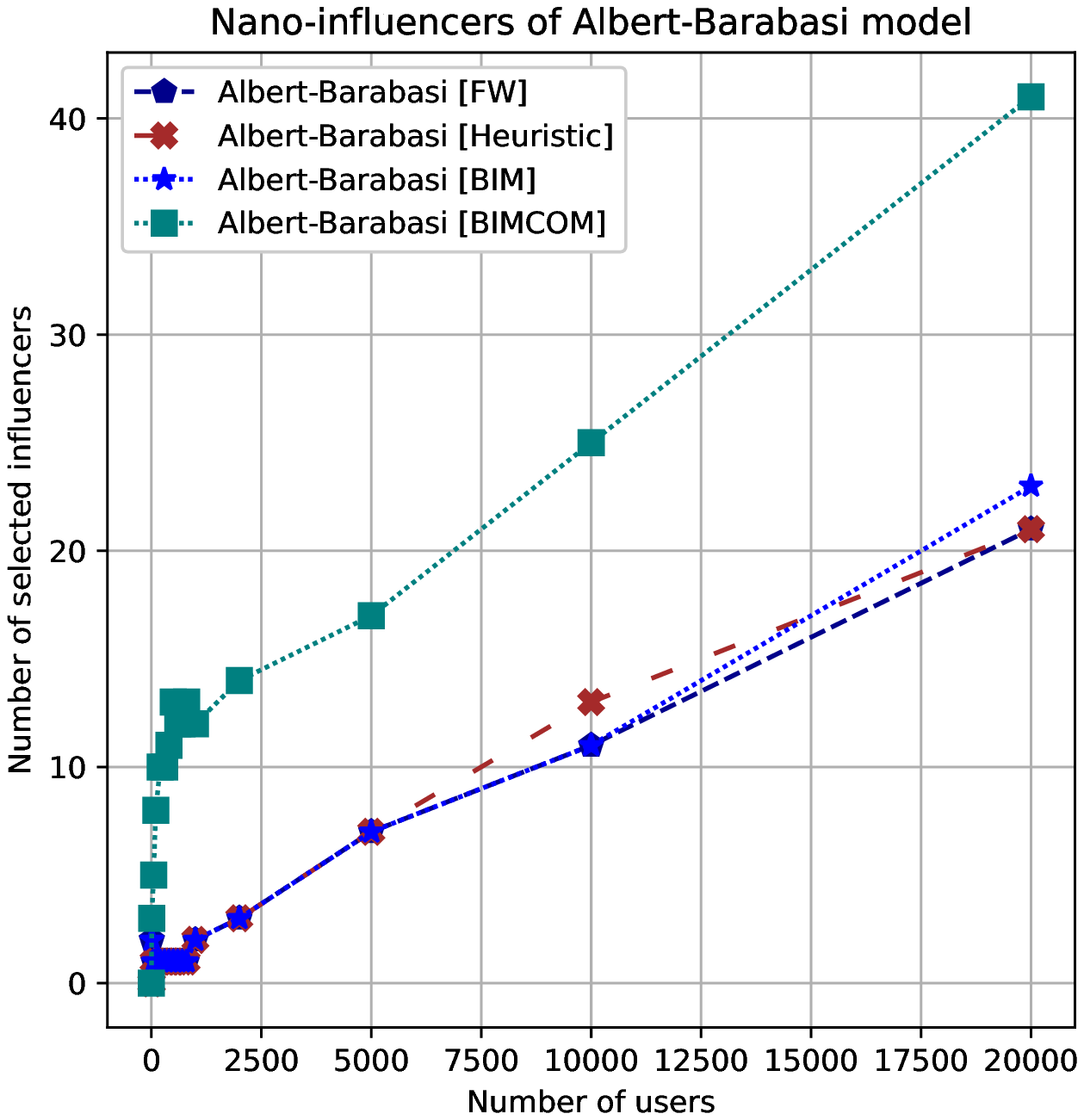}
     \end{subfigure}
     \hfill
     \begin{subfigure}[b]{0.2\textwidth}
         \centering
         \includegraphics[width=1.1\textwidth]{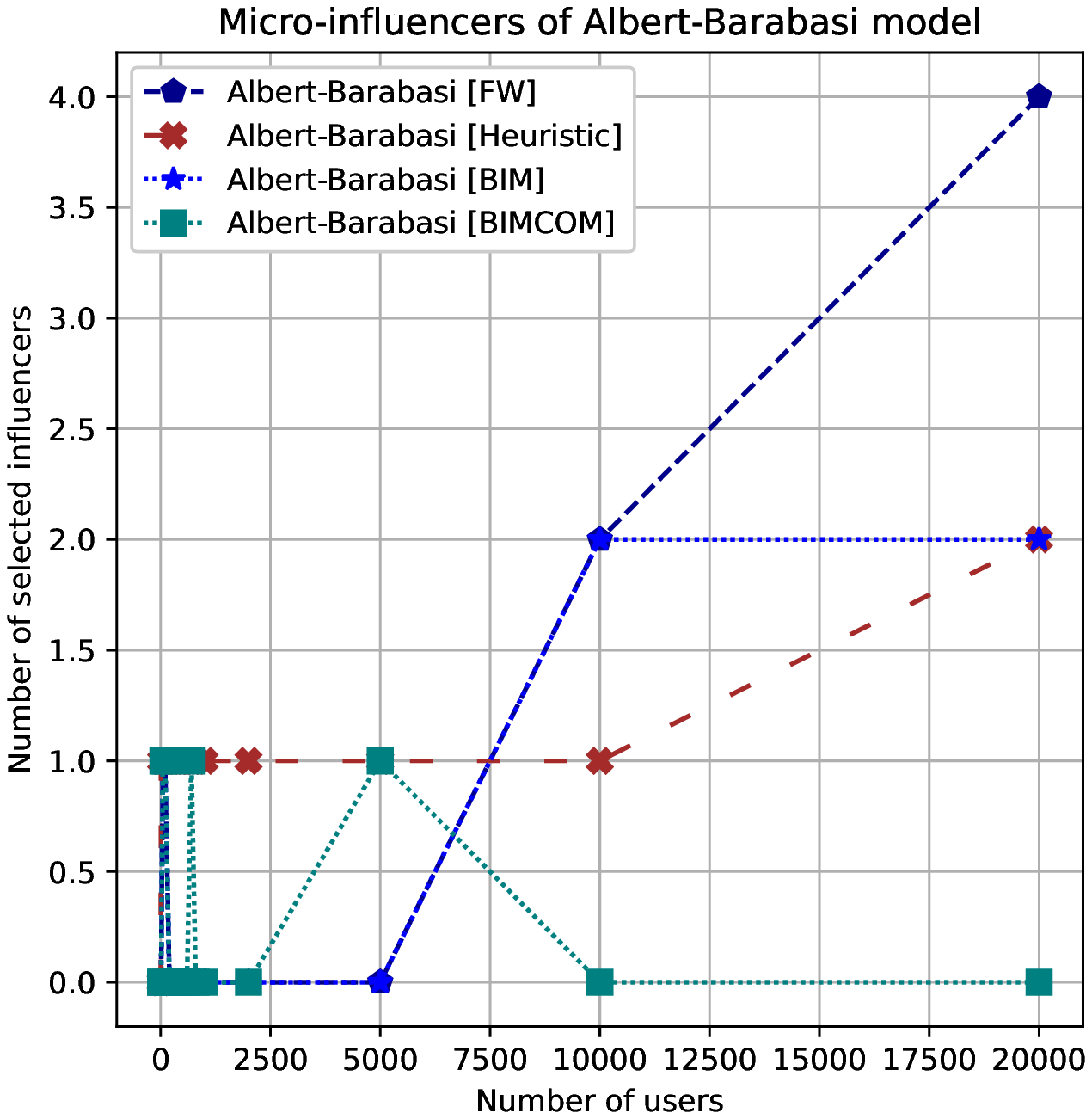}
     \end{subfigure}
     \hfill
     \begin{subfigure}[b]{0.2\textwidth}
         \centering
         \includegraphics[width=1.1\textwidth]{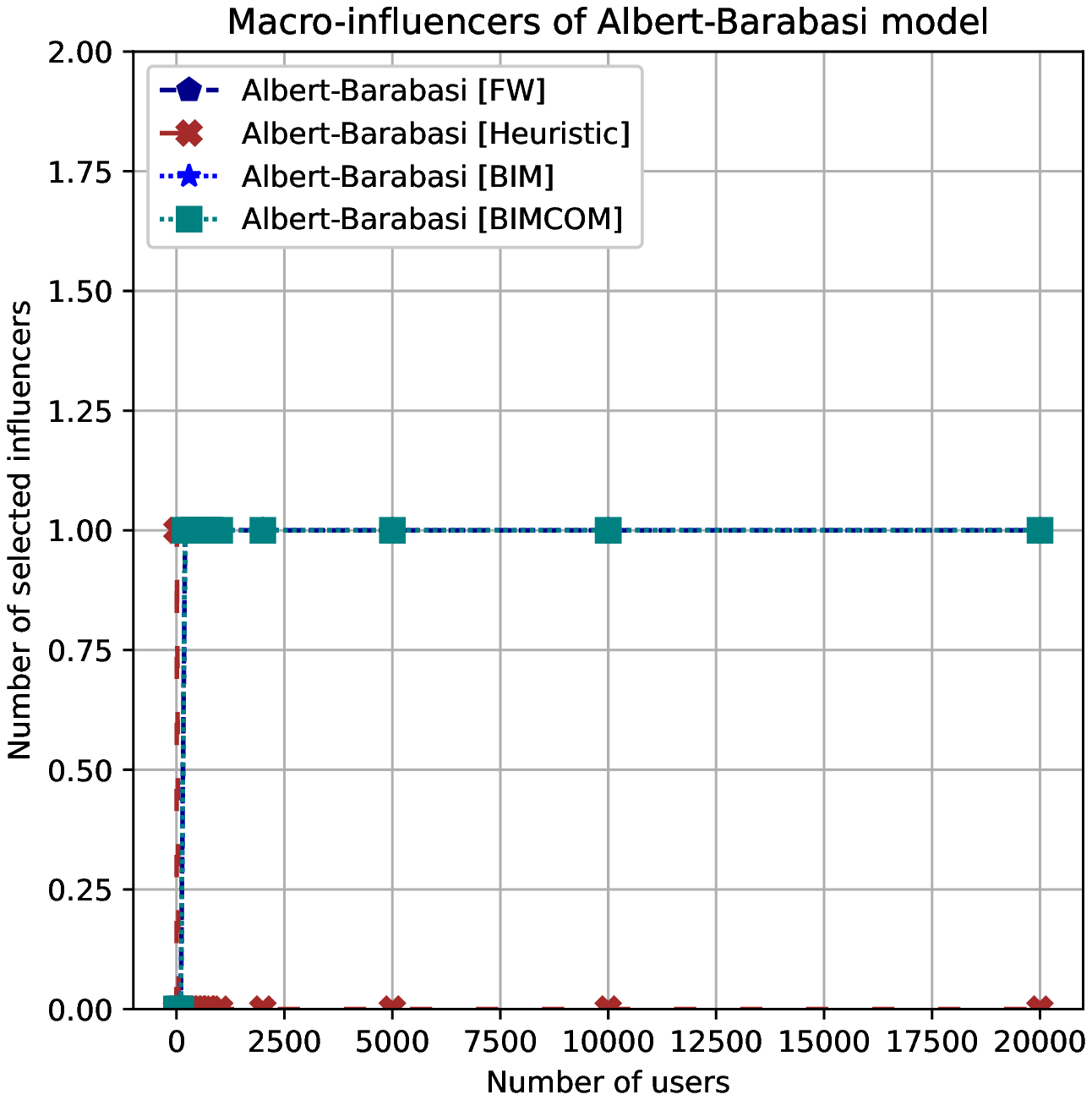}
     \end{subfigure}
     \hfill
     \begin{subfigure}[b]{0.2\textwidth}
         \centering
         \includegraphics[width=1.1\textwidth]{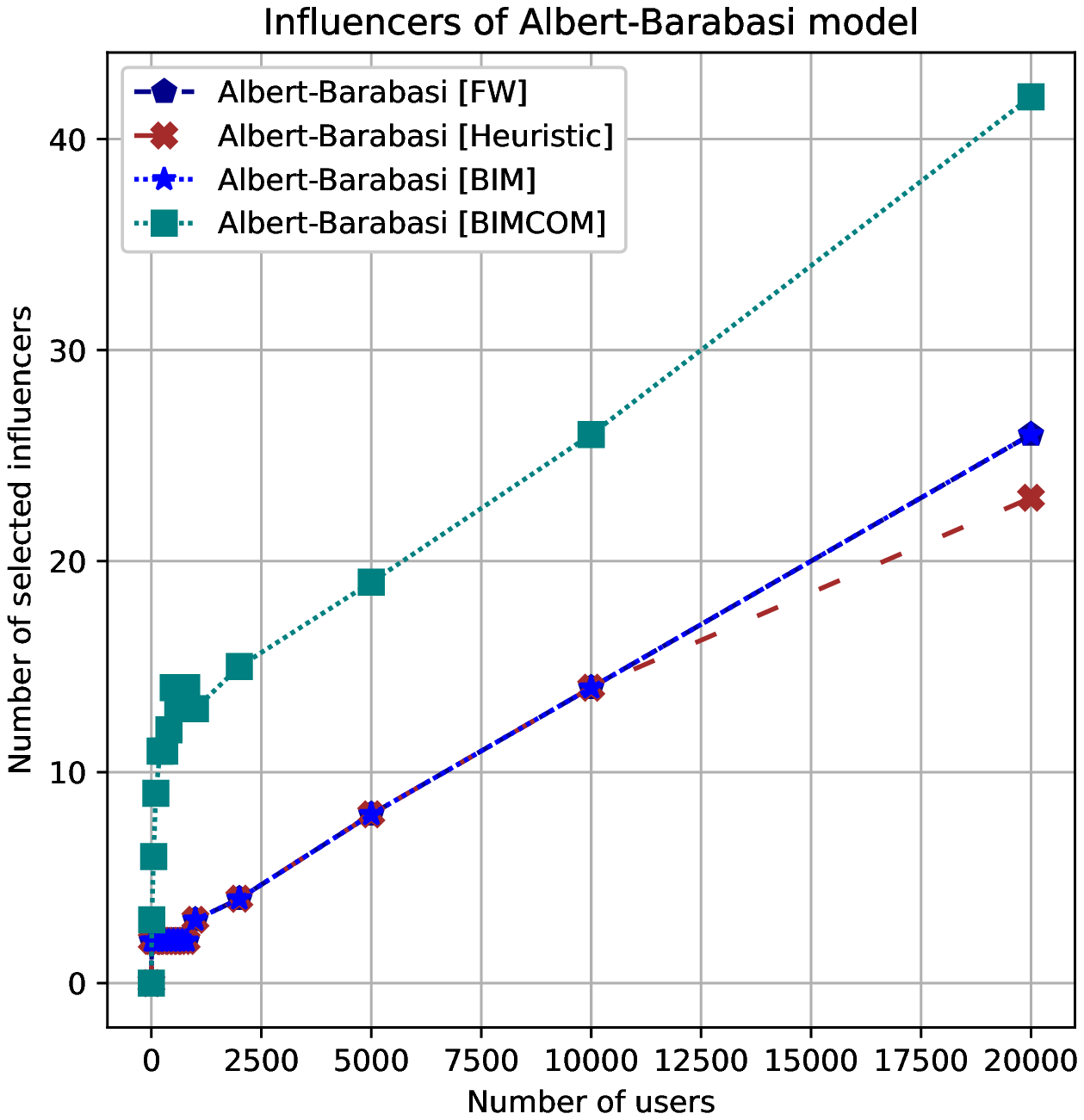}
     \end{subfigure}
        \caption{Number of selected influencers by category (i) Nano-, (ii) Micro-, (iii) Macro-, (iv) Total-, in the Albert-Barabasi model for the solution with Frank-Wolfe (FW), Budgeted Influence Maximization (BIM), the community-based approach for the Budgeted Influence Maximization (BIMCOM) and the rule of thumb (Heuristic).}
        \label{fig3}
\end{figure*}

We consider user sets 
$\mathcal{N}$ of size $N \leq 20,000$ 
within an hourly time window, and we investigate the following network structures:

\begin{enumerate}
    \item $\textit{Albert-Barabasi (\texttt{AB}) model:}$ An undirected graph generated by preferentially attaching new arriving nodes each with $A$ edges to existing ones in proportion to their degree. This is a scale-free network with a total of $A(N-A)$ edges.
\item $\textit{Erdos-Renyi (\texttt{ER}) model:}$ An undirected graph where each edge is included with probability $A(N-A)/\binom{N}{2}$, independently from every other edge. As in the \texttt{AB} model, \texttt{ER} has an expected number of $A(N-A)$ edges.
\end{enumerate}

For each network structure, we choose homogeneous posting rates, namely, $\lambda^{(n)}=\lambda$ [posts/time window]$, \forall n \in \mathcal{N}$, so that we can better investigate the effects of network structure and cost per post on the campaign. Values for the average impression ratios $\{ p^{(j)}_n \}_{n,j \in \mathcal{N}}$ 
are calculated analytically using the Markovian method introduced in \cite{a15} (see section \ref{sec:II}), by setting the re-posting rates as constants equal also to $\lambda$ for this model. We assume complete knowledge of the social graph, the posting rates and the impression ratios. For numerical studies, we select $A=4,\ \lambda=1$ to result in a sparse matrix of impressions as the network size increases, i.e. $D \ll_{N \rightarrow \infty} N^{2}$, where $D$ is the number of pairs with positive impressions.


We consider for the study a budget proportional to the network size $B=\frac{1}{100} N$ [EUR/time window]. Note that the above budget is normalized and it is invested in an hourly time window, so although this metric is low, the time window is also low. In addition, we will assume that the campaign is coordinated by the user $i=0$, where the number of followers will be different in each model.

As a next step, we determine the price per post $c_n$ charged to the advertiser $i$ by user $n$. On Twitter, it is a common and approximate market practice to consider the price per post of user $n$ as $2\frac{\textit{\#Followers}_n}{1000}$ [EUR/post] by \cite{RIn}. So, for our simulation purposes with small networks we omit the factor $\frac{1}{1000}$ in the price to get reasonable budget values. We consider no restrictions on user participation ratios, i.e. $r_n=1, \forall n \in \mathcal{N}$. For user utilities, we use the logarithmic function $U_j(x) =log(\delta x+1), \forall j \in \mathcal{N}\setminus \{i \}$ which quantifies the ROI of the advertising campaign, with a $\delta=1,000$ selected for simulation purposes.



Influencers are distinguished by their ability to disseminate posts through their followers. For the aim of our numerical simulation study, we will define the influence of a user $n$ based on their relative number of followers. So, we denote a user $n \in \mathcal{N}$ in a network as a:
\begin{itemize}
    \item \textit{Nano-influence}: If the number of his/her followers is up to the sixth decile of the degree distribution. 
    \item \textit{Micro-influencer}: If the number of his/her followers is greater than the sixth decile of the degree distribution, but not higher than the ninth decile.
    \item \textit{Macro-influencer}: If the number of his/her followers is not less than the ninth decile of the degree distribution. 
\end{itemize}

The advertiser in the Albert-Barabasi model and in the Erdos-Renyi model will most likely be a Nano-influencer or a Micro-influencer due to degree statistics. 
\smallskip
\subsubsection{Frank-Wolfe benefits}
The first point we want to illustrate is the benefit of our Frank-Wolfe algorithm compared to alternative optimisation approaches that we could have used. All these algorithms were programmed to adapt to~\ref{[BPO-G]}. We compare the solution of~\ref{[BPO-G]} using the Frank-Wolfe (FW) algorithm adaptation and the rule of thumb with two baselines of comparisons:
\begin{enumerate}
    \item Considering our model and the optimization problem~\ref{[BPO-G]} already established. In this case we use the Projected Subgradient method and the Mirror Descent, which are two highly used algorithms in the convex optimization bibliography.
    \item Considering a different approach of the problem with the Kempe's philosophy, which is commonly used in the influence maximization problem. In this case our Frank-Wolfe variation is compared against the budgeted influence maximization (BIM) approach \cite{a12}, and its variant the community-based approach to the budgeted influence maximization problem (BIMCOM) \cite{rec1}.
\end{enumerate}

In detail, the six compared algorithms are:
\begin{itemize}
    \item The Frank-Wolfe (FW) algorithm adaptation introduced in section~\ref{sec:III}.
    \item The Projected Subgradient (PS) method \cite{V2}; here we can not avoid the projection on the feasibility set, due to the restriction of $a_n\in[0,1].$
    \item The Mirror Descent (MD) method \cite{V1}; it adapts the PS method to the geometry of the~\ref{[BPO-G]} problem by the Bregman divergence associated to the function $\mathbf{a}\in \mathcal{M} \rightarrow \sum_{j \in \mathcal{N}\setminus \{i \}  | a_j \not = 0} a_j log(a_j).$
    \item The Budgeted Influence Maximization (BIM) approach \cite{a12}; it solves the BIM problem by a CELF implementation \cite{celf} of $1,000$ Monte Carlo simulations under the assumption of an independent cascade model, with a probability of influence being propagated between any two users equal to $p=\frac{\sum_{n,j \in \mathcal{N}}  (p^{(j)}_n)^{1/k}}{N^2}$ with $k$ the average shortest path between any two users.
    \item A community-based approach for the Budgeted Influence Maximization (BIMCOM) problem \cite{rec1}; This approach consists of four steps: a community detection to understand the structure of the network, a budget distribution to divide the budget among the communities, seed selection for influence maximization and finally, budget transfer, in which unutilized budget of one community is transferred to another community. Similarly as in BIM we considered $1,000$ Monte Carlo simulations under the assumption of an independent cascade model, with a probability of influence propagation equal to $p$.
    \item The rule of thumb we presented in Section~\ref{rule}.
\end{itemize}  

The stopping criterion in the first three algorithms (FW, PS and MD) is when the number of iterations reaches up to $20$, or when the theoretical optimality gap is less than $0.1$. 

We compare the six methods across the different network structures on the following performance metrics for \texttt{ER} and \texttt{AB}: (i)-(ii) the solution optimum, and (iii)-(iv) the runtime. 
The plots in Fig.~\ref{fig2} illustrate how the above metrics vary as we increase the size of the network across the different network structures and for each algorithm executed. Note here that the optimum curve does not exhibit diminishing returns as shown in (i)-(ii) Fig.~\ref{fig2}, because the budget is increasing proportionally to the network size.

All three optimisation algorithms (FW, PS and MD) find the same solution optimum as we show in (i)-(ii) Fig.~\ref{fig2}. An almost optimal solution is found for BIM and for the rule of thumb in \texttt{ER}, and a not so good approximation for the solution optimum in \texttt{AB}. The Table \ref{table:3} shows the relative errors of the optimum found between FW, BIM and the rule of thumb across different network structures for two $\delta$-values and two network sizes. We observe that a decrease in the network size or an increase in $\delta$ reduces the relative errors of the optimum between FW, BIM and the rule of thumb, which is expected given the selection of the utility functions and the increments in the optimum shown in (i)-(ii) Fig.~\ref{fig2}. Therefore, our empirical results found in Table \ref{table:3} verify the results found by \cite{nature}. In addition, a worse performance in BIMCOM compared to BIM in the optimum is observed, since that BIMCOM breaks down the network into communities and loses information from the interconnection of such communities. So, the performance in the optimum can be worse in BIMCOM than in BIM.


Considering the execution time, we observe in (iii)-(iv) Fig.~\ref{fig2} that the rule of thumb is the fastest tested algorithm regardless of the network structure and the network size, so we have a significant trade off in runtime and solution optimum. Note that, Frank-Wolfe which finds the exact optimum is faster than MD, PS, and BIM for all network structures and network sizes. As expected, we observe a continuous increase in the runtime of BIM except when a disconnected network is encountered, which speeds up the algorithm and is represented as a bump in (iii)-(iv) Fig.~\ref{fig2}. So, the execution time of BIM is longer than FW, which we know converges from theoretical results. Besides, we observe that BIMCOM is faster than BIM since that by breaking down the network into smaller communities speeds up each BIM sub-process, when in fact BIM considers the whole network, which takes more time. Hence we have a trade off in runtime and distance from optimum with BIM and BIMCOM. As a side result, MD is faster than PS regardless the network structure and the network size. This is natural since MD takes advantage of the geometry of the feasible set $\mathcal{M}$ in the equation~(\ref{eqfeas}). 


We can generalise our observation for BIM and BIMCOM also for other algorithms which use some arbitrary information diffusion model in OSPs (e.g. the Linear Threshold model or the Independent Cascade model); we expect that such model-based algorithms will underestimate the optimum because these cannot accurately mimic the true propagation of influence among users. However, this is avoided by our suggested algorithm by incorporating measured influence data in the objective function, this is why it exhibits the highest performance.

\begin{table}[t!]
\resizebox{\columnwidth}{!}{
    \centering
\begin{tabular}{||l | c c c c||} 
 \hline
  Network \& Algorithm   & $(10K,5)$ & $(20K,5)$ & $(10K,1000)$ & $(20K,1000)$ \\ [0.5ex] 
 \hline
 Erdos-Renyi, FW vs BIM  & 15.456\% & 15.836\% & 13.613\% & 13.957\% \\ 
 \hline
  Erdos-Renyi, FW vs Heuristic & 15.995\% & 16.171\% & 13.644\% & 14.247\% \\ [1ex] 
 \hline
 Albert-Barabasi, FW vs BIM & 31.835\% & 42.835\% & 18.333\% & 29.627\% \\
  \hline
 Albert-Barabasi, FW vs Heuristic & 51.072\% & 76.723\% & 27.706\% & 49.450\%   \\
 \hline
\end{tabular}}
    \caption{Relative improvement in optimum of FW compared to other methods, for graph sizes $N\in\{10K,20K\}$ and utility parameter $\delta\in\{5,1000\}$\\}
    \label{table:3}
    \vspace{-1cm}
\end{table}

\smallskip
\subsubsection{Campaign-related results}
We compare the selected number of Nano-, Micro-, and Macro-influencers from FW, BIM, BIMCOM and the rule of thumb (Heuristic) for each network structure.


The plots in Fig.~\ref{fig3} show the distribution of the number of selected influencers by category in \texttt{AB}. We observe that both BIM and especially the rule of thumb (Heuristic) follow the distribution of the optimal solution by the Frank-Wolfe (FW) algorithm. 
In the Albert-Barabasi network having heavy-tail degree distribution, we can see in Fig.~\ref{fig3} an increasing preference for Micro-influencers and Macro-influencers when the size of the network increases regardless the algorithm selected. 
Interestingly, the rule of thumb selects the fewest number of influencers regardless the network size between the tested algorithms, which is shown in Fig.~\ref{fig3}. On the other hand, BIMCOM selects more Nano-influencers than BIM, by splitting the graph into smaller communities. As a consequence of the above, BIMCOM selects a larger number of influencers compared to the other methods. When the network is \texttt{ER}, the algorithm selects in majority Nano-influencers, which is expected due to the small degree variance of \texttt{ER} graphs. Finally, among both network structures, \texttt{AB} selects the minimum total number of influencers, which is expected since this model has a heavy-tail degree distribution. Hence, our model captures the underlying structure of the network in a good manner. 



\begin{figure*}[!t]
     \centering
     \begin{subfigure}[b]{0.3\textwidth}
         \centering
         \includegraphics[width=1\textwidth]{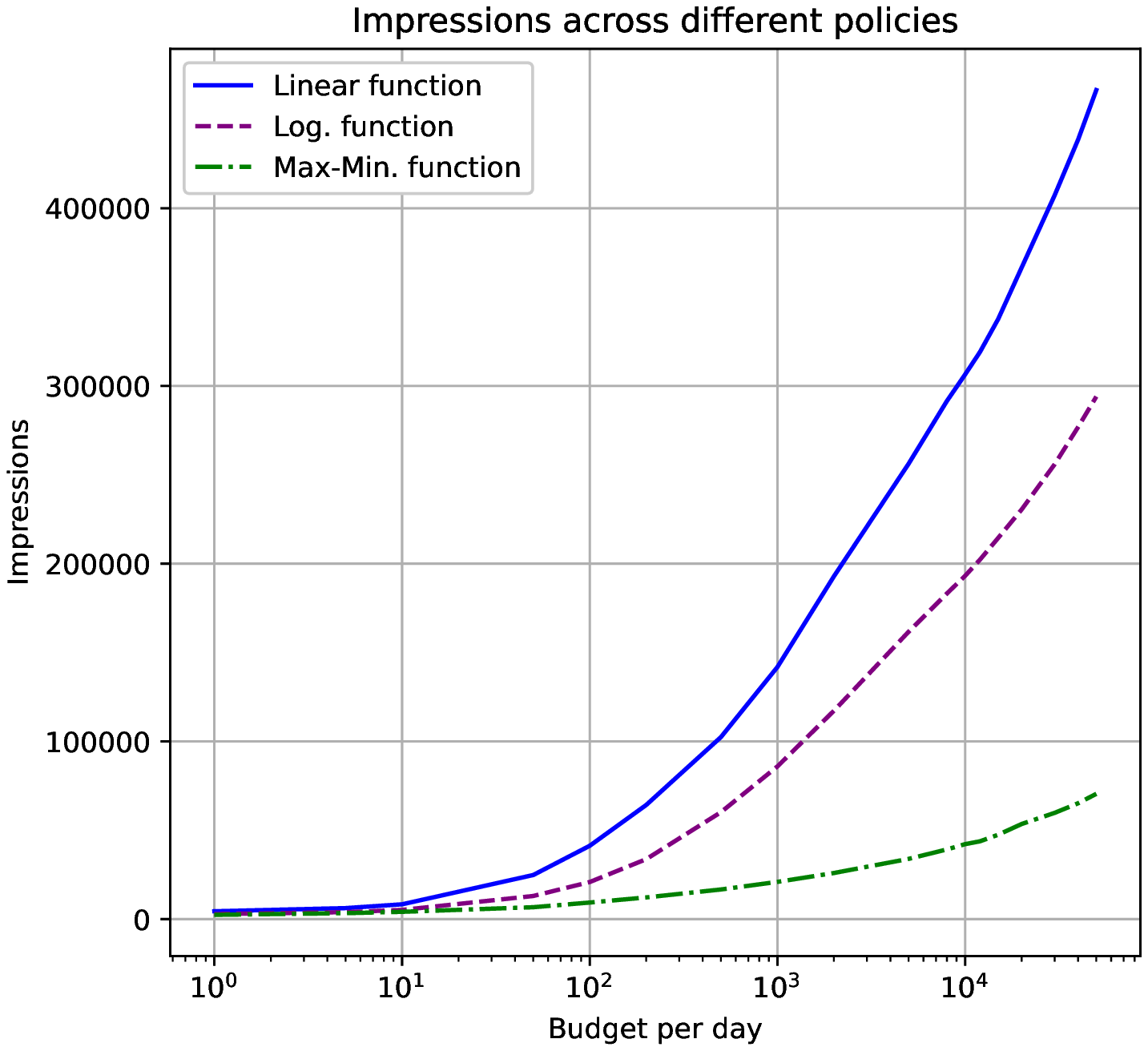}
     \end{subfigure}
     \hfill
     \begin{subfigure}[b]{0.3\textwidth}
         \centering
         \includegraphics[width=1\textwidth]{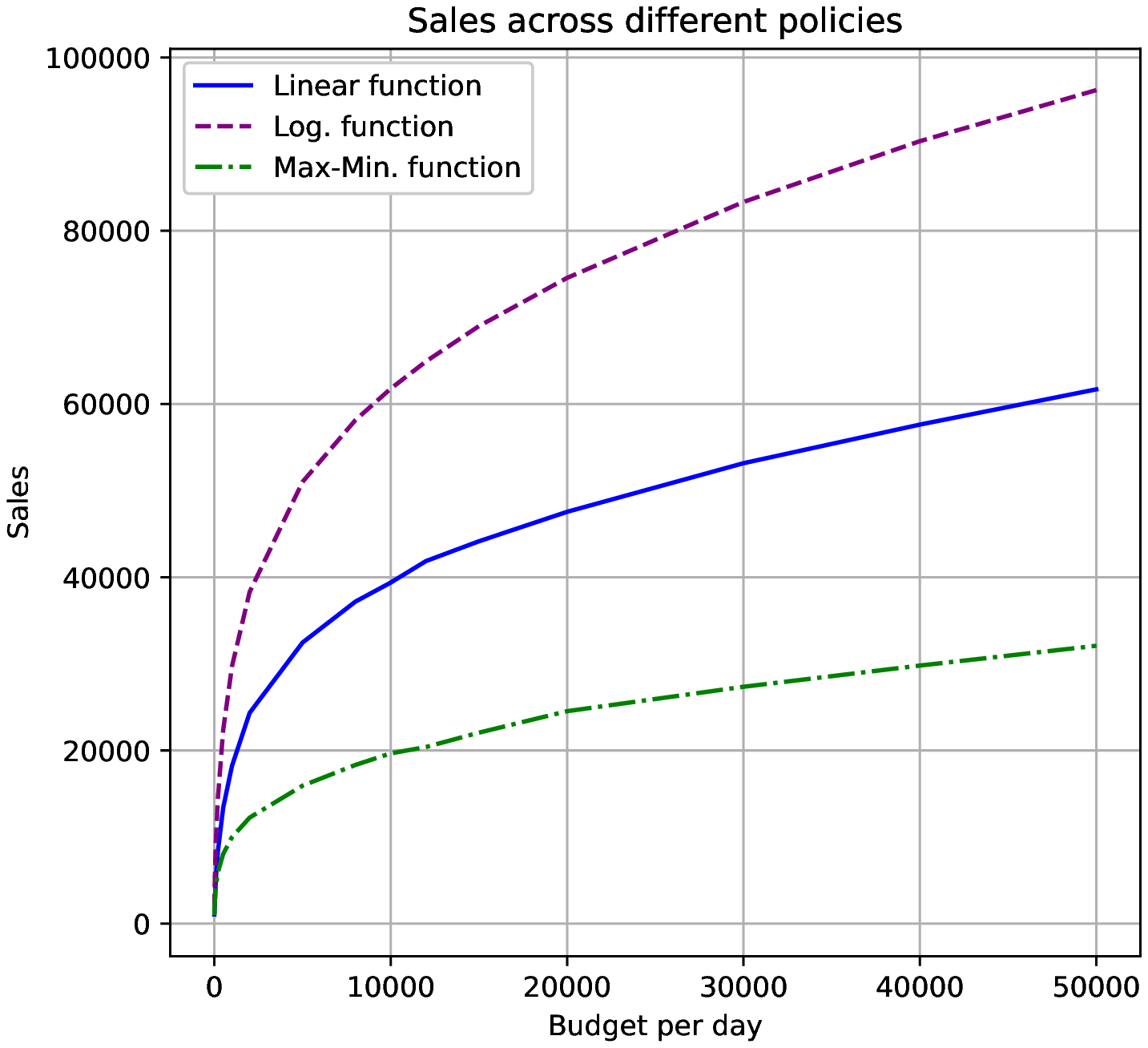}
     \end{subfigure}
     \hfill
     \begin{subfigure}[b]{0.3\textwidth}
         \centering
         \includegraphics[width=1\textwidth]{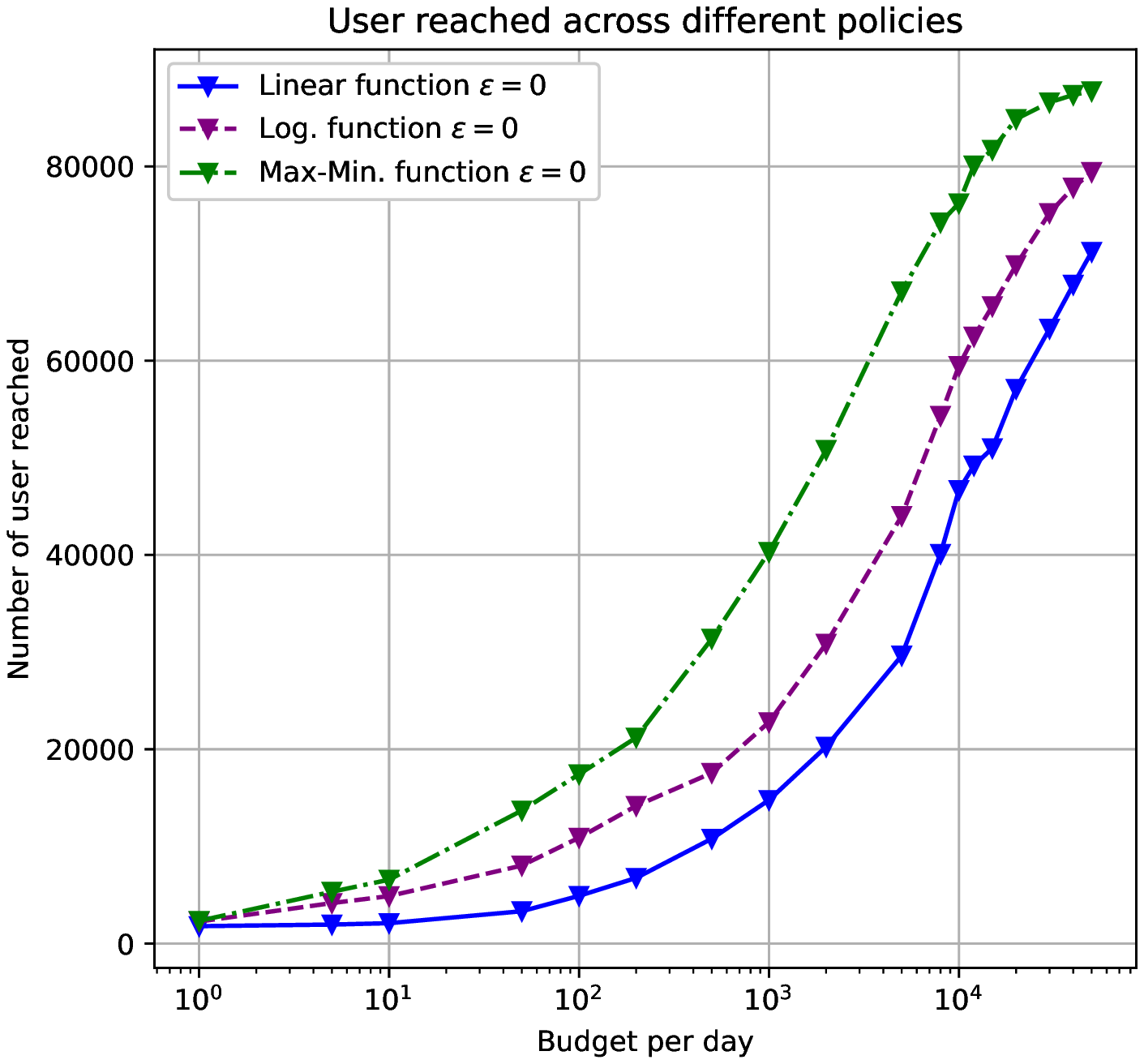}
     \end{subfigure}
    \bigskip
     \centering
     \begin{subfigure}[b]{0.3\textwidth}
         \centering
         \includegraphics[width=1\textwidth]{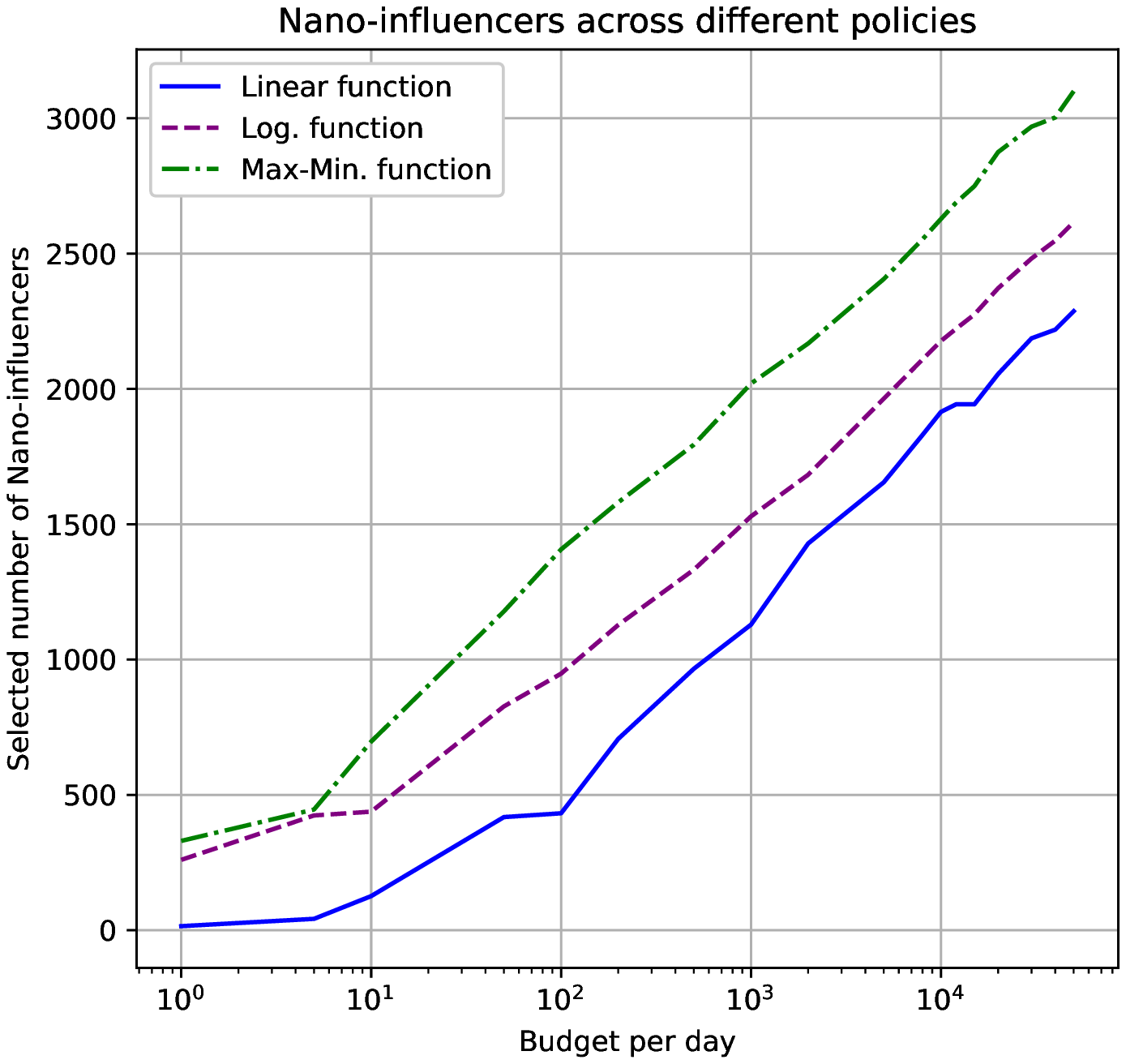}
     \end{subfigure}
     \hfill
     \begin{subfigure}[b]{0.3\textwidth}
         \centering
         \includegraphics[width=1\textwidth]{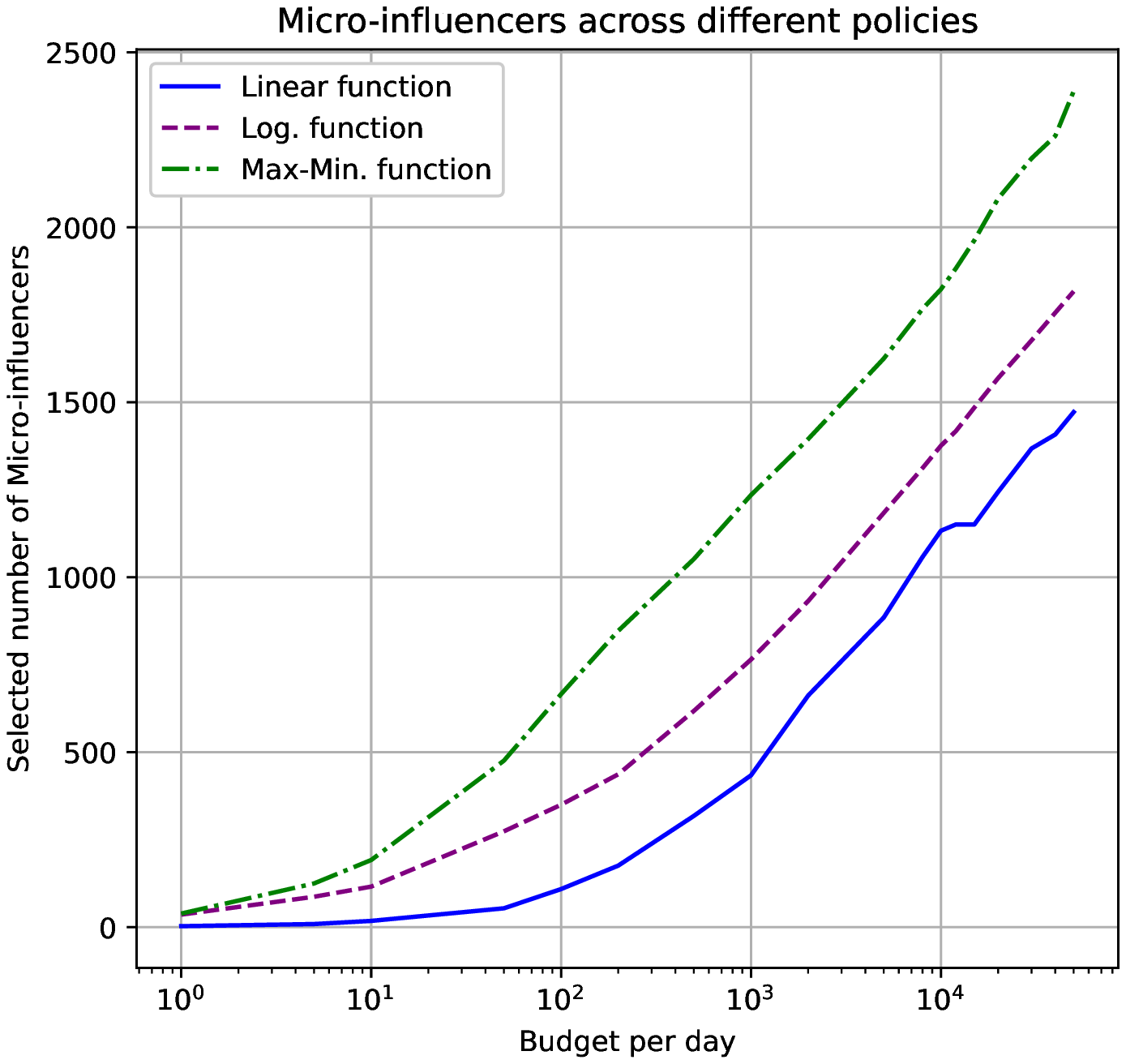}
     \end{subfigure}
     \hfill
     \begin{subfigure}[b]{0.3\textwidth}
         \centering
         \includegraphics[width=1\textwidth]{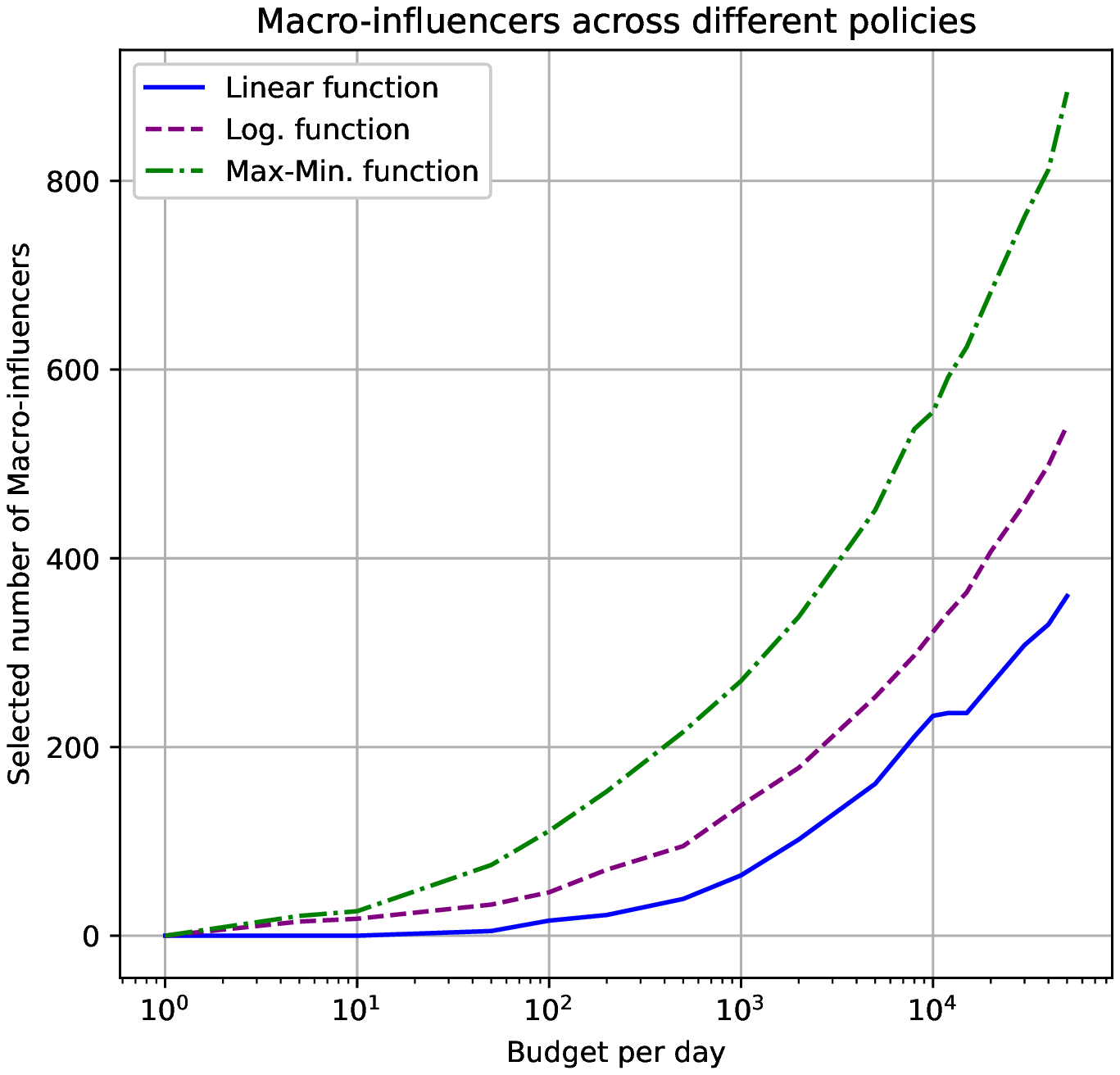}
     \end{subfigure}
     \caption{Twitter dataset: Metrics across different campaign policies and number of selected influencers across different campaign policies.}
    \label{fig4}
\end{figure*}

\subsection{Numerical evaluations on a real Twitter data trace}
\label{sec:V-B}

The aim of this subsection is to evaluate the performance of our Algorithm~\ref{algo2}, for various campaign objectives introduced in section~\ref{sec:III} by using information from a real large Twitter data trace \cite{a54}. This database represents the activity of users on Twitter during the 2018 Russian elections. 

In particular for our purposes, we use a 4-uple per post with the following information obtained from this database: $[\texttt{TweetID,\ TimeStamp,\ UserID,\ RetweetID}]$. 
Each participating user and Tweet have a unique associated UserID and TweetID respectively. RetweetID represents the TweetID which was retweeted (or $-1$ if it is a self-post) and TimeStamp is the time that the Tweet was (re)-posted. The entire database spans $57$ days and involves $181,621$ different $UserID$s. Moreover, there is an average of $3.71$ posts, an average of $7$ re-posts per user and we find $87,987$ users who have re-posted (shared a post) at least once. These users can be potentially reached by any advertising campaign, since they share content.

From the dataset, we derive the empirical post and re-post rate for every user $\{ \lambda ^ {(n)} \}_{n \in \mathcal{N}}$ and $\{ \mu ^ {(n)} \}_{n \in \mathcal{N}}$ respectively. We can further infer a friendship graph using the relationships of retweets (RetweetID), by drawing a directed edge from leader to follower, each time a user retweets something. We call this a "star" shaped graph due to its structure: it contains $181,621$ nodes, $517,421$ edges with a mean degree of $5.70$ followers per user. Among the users, $167,646$ users lack of followers and only $13,975$ users have followers.

The Twitter data trace we are working with \cite{a54} contains in each line the 4-uple we mention above, and no extra information is available; consequently we can only reconstruct a star friendship-graph based on the available trace information and the involved users appearing inside by either posting or reposting. The extra information how many followers each user actually has inside the complete Twitter graph is not available to us. We can only extract how many users follow each user inside the graph derived by the trace-log. Hence, it is necessary to redefine what is Nano-, Micro-, and Macro-influencer based only on the sampled database. For this purpose, we classify the $13,975$ potential influencers ($\lambda^{(n)}>0$) into $3$ categories: $8,615$ users have $1-3$ followers and are potentially Nano-influencers; $3,986$ have $4-34$ followers and are potentially Micro-influencers; and $1,374$ have more than $34$ followers and are potentially Macro-influencers. Such partition follows the definition given in Section~\ref{sec:V-A} based on the percentages (deciles) of the degree distribution of followers given the available (but not the complete Twitter) graph. Using the derived social star-graph through the trace-log and the posting and re-posting rates, the average impression ratios $\{ p^{(j)}_n \}_{n,j \in \mathcal{N}}$ 
can be estimated by the Markovian method introduced in \cite{a15} (see Section II.A). By definition, the engagements are the shared impressions during the $57$ days. 

As a next step, we need to determine the price per post $c_n$ charged to the advertiser user $i$ by user $n$. In this case, we will use the common and approximate market practice on Twitter described in~\ref{sec:V-A} normalized through the number of users in the database. Observe that our database is of the order of $10^5$ users and Twitter is of the order of $10^8$, so we will assume a normalization constant in the number of followers of $10^3$, so our price per post of user $n$ to consider is $2 \textit{\#Followers}_n$ [EUR/post].

For the evaluations we will consider no restrictions on user participation ratios ($r_n=1, \forall n \in \mathcal{N}$) in the absence of information. Finally, we select as advertiser the user with $\text{UserID}=2513730044$, who has $15\textit{\#Followers}$. This user is potentially a Micro-influencer.

\begin{table}[t!]
\resizebox{\columnwidth}{!}{
    \centering
\begin{tabular}{||l | c c ||} 
 \hline
   Network \& Algorithm   & FW vs BIM & FW vs Heuristic \\ [0.5ex] 
 \hline
  Twitter dataset & 23.645\% & 33.533\% \\ 
 \hline
\end{tabular}}
    \caption{Relative improvement in sales of FW compared to other methods for a budget of $10,000$ [EUR/day] with $\delta=10$ in a Log function.\\}
    \label{table:4}
    \vspace{-1cm}
\end{table}

We proceed to solve and to find the optimal solutions $\mathbf{a}^{*}_L, \mathbf{a}^{*}_{Log}$ and $\mathbf{a}^{*}_M$ of~\ref{[BPO-G]} according to three functions respectively: Linear function, Logarithmic function and Max-min function. The stopping criterion of our Algorithm~\ref{algo2} is when the number of iterations reaches a maximum equal to $30$, or when the optimum gap defined by equation~(\ref{ceq26}) is less than $0.1$. Using these solutions $\mathbf{a}^{*} \in \{\mathbf{a}^{*}_L, \mathbf{a}^{*}_{Log}, \mathbf{a}^{*}_M \}$, we evaluate the metrics: 

$\bullet$ \textit{Total number of Impressions}: $\delta \ \sum_{j \in \mathcal{N}\setminus \{i \} } \omega^{(j)}(\mathbf{a}^{*})$.

$\bullet$ \textit{Total Sales}: $\sum_{j \in \mathcal{N}\setminus \{i \} } log(\delta \ \omega^{(j)}(\mathbf{a}^{*})+1)$.

$\bullet$ \textit{Total Reach}: $\sum_{j \in \mathcal{N}\setminus \{i \} } I_{\omega^{(j)}(\mathbf{a}^{*})>\epsilon},$ $\epsilon$ is a threshold and denotes when a user has been reached by the campaign. So, we will evaluate $\epsilon=0$ as a representative of the reach. 

$\bullet$ \textit{Selected number of Nano-, Micro-, and Macro-influencers}: The number of users with $a_j^{*}\not = 0$. Specifically: those up to $3$ followers are Nano- (fewer followers than $60\%$ of users in our database), those with $4-34$ followers are Micro- (no more followers than $60\%$ of users but fewer than $90\%$ in our database), and those with more than $34$ followers are Macro- (more followers than $90\%$ of users in our database).

A constant $\delta=10$ in the above is selected to carry out the numerical evaluation on the real Twitter data trace. Let us first observe that similarly as in Section~\ref{sec:V-A}, the results of our comparisons are extendable over the real Twitter data trace, on the relative improvement in optimum of the Frank-Wolfe algorithm adaptation compared to other methods and as shown in the Table \ref{table:4}. We note that the Frank-Wolfe algorithm adaptation gives us the best performance in the example case of sales for a fixed budget of $10,000$ [EUR/day] with the Log utility function. Therefore, our empirical results found in Table \ref{table:4} verify the results found by \cite{nature}. The plots in Fig.~\ref{fig4} illustrates how the above metrics change with increasing monetary budget per day, for each of the three different campaign objectives (Linear, Logarithmic and Max-min). We observe from Fig.~\ref{fig4} the following:

\textit{Linear objective} - This campaign gives the highest impressions performance. It gives moderate sales for any budget, but has the worst reach performance. It selects the least number of influencers in all categories, for any budget. 

\textit{Logarithmic objective} - This campaign has the highest sales performance. Also, it has a high impressions for a budget $>1000$ [EUR/day] and a moderate reach performance. It selects more influencers than the linear, in all categories.

\textit{Max-min objective} - This campaign gives the highest audience reach for any given budget, but performs bad in sales and impressions. In fact, for a budget $>40K$ [EUR/day] the campaign can reach all possible users. 

For all three objectives, the optimal policy selects mostly Nano- and Micro-influencers in low budgets. Macro-influencers are selected for larger budgets. This is follows closely how selection is done in practice. In addition, we observe that the sales performance per day behaves as a curve with diminishing returns for all budgets and the three campaign objectives, which is natural and expected.


\subsection{Multi-platform evaluation}

We now study the multi-platform model introduced in section~\ref{sec:IV}, to evaluate how the optimal solution splits the budget and consequently the ROI between $L=2$ networks, one synthetic and one from the Twitter data trace \cite{a54}.

From the Twitter dataset \cite{a54} we inferred in~\ref{sec:V-B} the "star" shaped graph with size $N_{R}=181,621$ users. 
As second graph we use an Albert-Barabasi network with size $N=5,000$ users. 
For the star graph we use the empirical posting and re-posting rates per user to derive the average impression ratios calculated via the Markovian method in~\ref{sec:V-B}. Similar process is followed for the AB model, albeit posting and re-posting rates are here chosen equal for all users $\lambda_{AB,n}=\lambda_{AB}=1$ [posts/day]. The choice of the imbalance between the size of the two networks is done on purpose to emphasize results.



The budget for both platforms is equal to $B=1,000$ [EUR/day]. In addition, we assume that the campaign is coordinated by the advertiser with ID in AB platform $i_{AB}=0$, and in Twitter  $\text{UserID}=i_{R}=2513730044$ and has $15\textit{\#Followers}$ in the second. 
For user utilities in both platforms, we consider the logarithmic function which quantifies the ROI of the advertising campaign. 

In our study, we want to examine how the budget and consequently ROI is split between the two platforms as the price per post ratio varies. So, we consider in the Twitter Star-graph, a price per post $c_{R,n}=c_R=1,  \forall n \in \mathcal{N}_{R}$ charged by every user. In the Albert-Barabasi model the price per post $c_{AB,n}=c_{AB}$ charged is equal among all users, but varies. Hence, the price-per-post-ratio $\frac{c_{AB}}{c_{R}}$ over both platforms varies. 

To solve the multi-platform model, we need to determine the set of non-negative constants $\{ \sigma_{l} \}_{l=AB,R}$ that maps the potential of a user in the platform $l$, to the respective ROI of the advertiser. For this case we will consider that the relative importance of the platform $l$ is proportionally linked to its price per post, so we will consider that $\sigma_{AB}=\frac{c_{AB}}{c_{AB}+c_{R}}, \sigma_{R}=\frac{c_{R}}{c_{AB}+c_{R}}$. 

We proceed to solve and to find the optimal solutions $\mathbf{a}_{x}^{*}$ of the optimization problem~\ref{[M-BPO]} by the elements of both platforms defined above and varying the price per post ratio $x:=\frac{c_{AB,n}}{c_{R,m}}>0, \forall n \in \mathcal{N}_{AB}, m \in \mathcal{N}_{R}$. 

\begin{figure}[!t]
    \includegraphics[width=\linewidth]{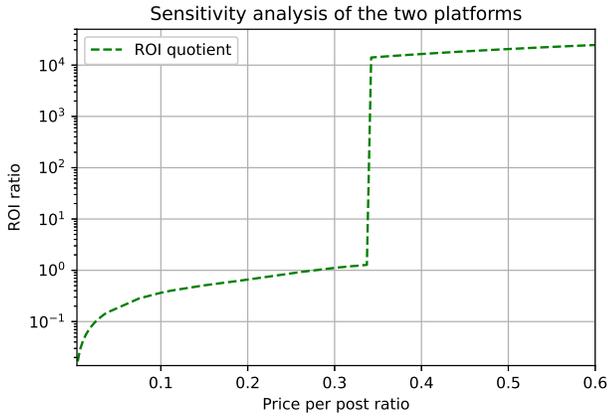}
    \caption{Optimal ROI split between two networks, for varying price ratio.}
    \label{fig5}
\end{figure}

The plot in Fig.~\ref{fig5} illustrates how the ROI ratio $\frac{\sum_{j \in \mathcal{N}_{R}\setminus \{2513730044 \}} \sigma_{R} log(\omega_{R}^{(j)}(\mathbf{a}_{x}^{*})+1)}{\sum_{j \in \mathcal{N}_{AB}\setminus \{i_{AB} \}} \sigma_{AB} log(\omega_{AB}^{(j)}(\mathbf{a}_{x}^{*})+1)}$ vary as we increase the price per post ratio $x$.

We observe in Fig.~\ref{fig5} that when $x$ is low, or else the $c_{AB}$ is low relative to $c_R$, then the solution tends to favorise the Albert-Barabasi model in budget allocation. The previous behaviour is compensated for the low $\sigma_{AB}$ (proportional to $c_{AB}$) and as a consequence, the ROI ratio is low since that virtually there is no selection of influencers in the Twitter graph. The trend described above continues up to a certain price ratio $x$ beyond which the trend reverses, and the ROI ratio starts increasing due to a high selection of influencers in the Twitter graph and starvation of influencer selection in the Albert-Barabasi graph. This happens, regardless of the marginal increase in $\sigma_{AB}$ as we can see in the Fig.~\ref{fig5}. As $x$ further increases, the growth in the ROI-ratio will moderate but will continue as long as there are posts to buy, albeit at a slower pace. This example shows how our model can nicely adapt to the multi-platform case, outputting reasonable results. 




\section{Conclusions} 
\label{sec:VI}

In this paper, we have presented an original continuous
formulation of the budgeted campaign orchestration problem to maximize some impact metric, e.g. the number of impressions, the sales, or the audience reach. We have derived a convex program, and further proposed an efficient iterative algorithm that converges to the global optimum efficiently. The iterative solution is based on the Frank-Wolfe algorithm, it has low computational complexity and high speed of convergence. We tested it against several algorithms from the optimization literature and found that our Algorithm~\ref{algo2} based on the Frank-Wolfe method has a superior performance in execution time and memory, and it scales well for problems with very large numbers (millions) of social users. Besides, for more practicality we have extracted a rule of thumb from our solution which although not optimal actually performs reasonably well for any campaign objective, offering a reasonable trade-off between runtime and optimality.

Generalisations of our continuous formulation of the budgeted campaign orchestration problem to Multi-platform and multi-content are also considered. In these generalisations, users contribute to different extents to an advertising campaign depending on the type of platform (e.g. Twitter, Facebook, Instagram) and the type of post (e.g. text, image, video) for the promotion of a target product. Our low-complexity fast algorithm based on the Frank-Wolfe algorithm naturally extends to such Multi-platform and multi-content instances. 

As future step, we can assume either that platform and user data (impressions and posting rates) are noisy, or that they evolve over time, and we should design an influencer campaign that optimally follows such fluctuations.

\ifCLASSOPTIONcaptionsoff
  \newpage
\fi



%

%

\begin{IEEEbiography}[{\includegraphics[width=1in,height=1.25in,clip,keepaspectratio]{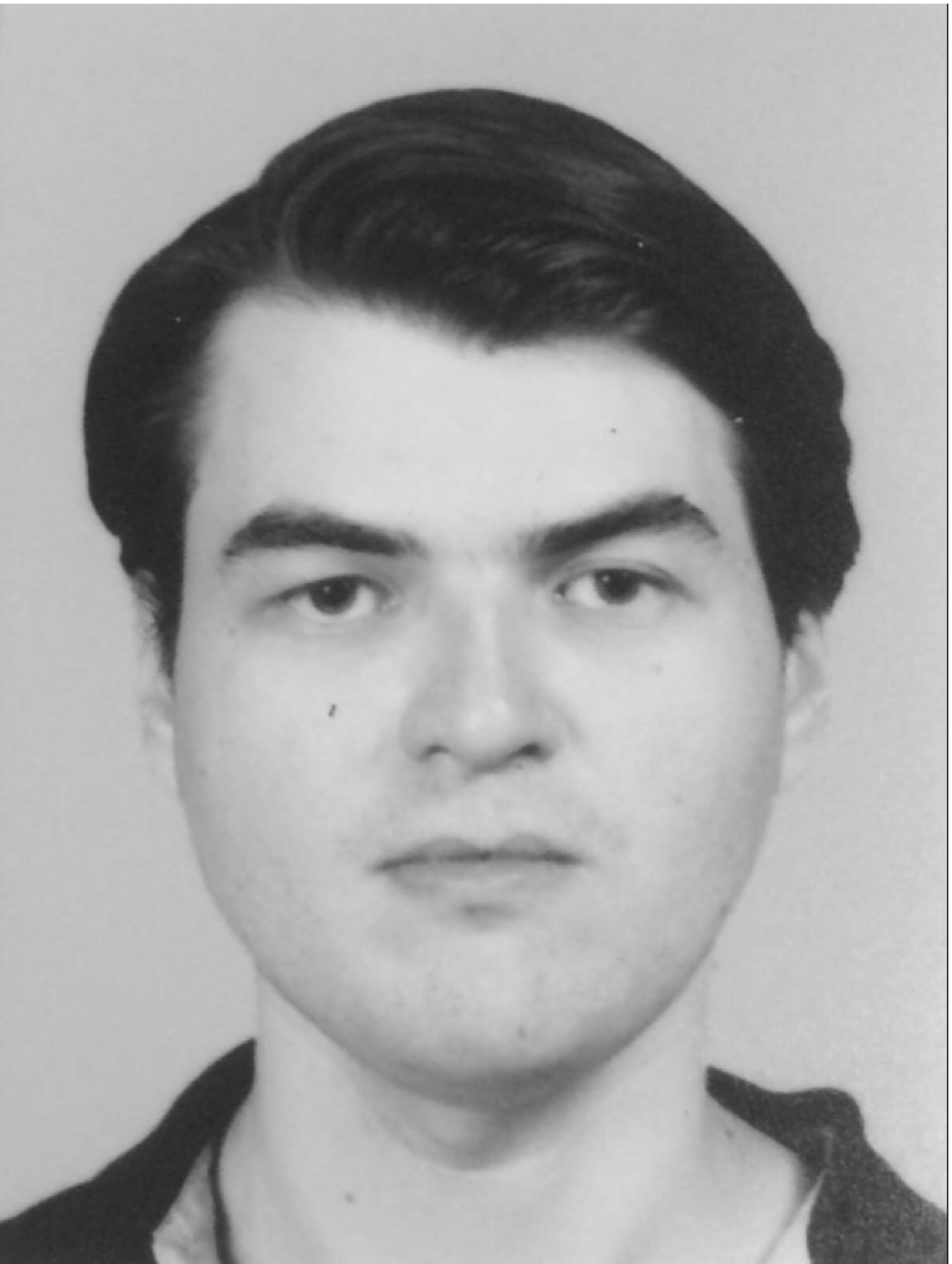}}]{Ricardo L\'opez-Dawn} (Member, IEEE)
received the B.S. degree in mathematics from the university of Guanajuato, Mexico,
in 2017 and the M.S. degree in probability and statistics from
center of research in mathematics, Mexico, in 2019. He is currently working toward the Ph.D.
degree with the LIP6-CNRS, Sorbonne Université, Paris,
France. His research interests are in mathematical and stochastic modeling, data analysis, optimization, and social networks.

\end{IEEEbiography}

\begin{IEEEbiography}[{\includegraphics[width=1in,height=1.25in,clip,keepaspectratio]{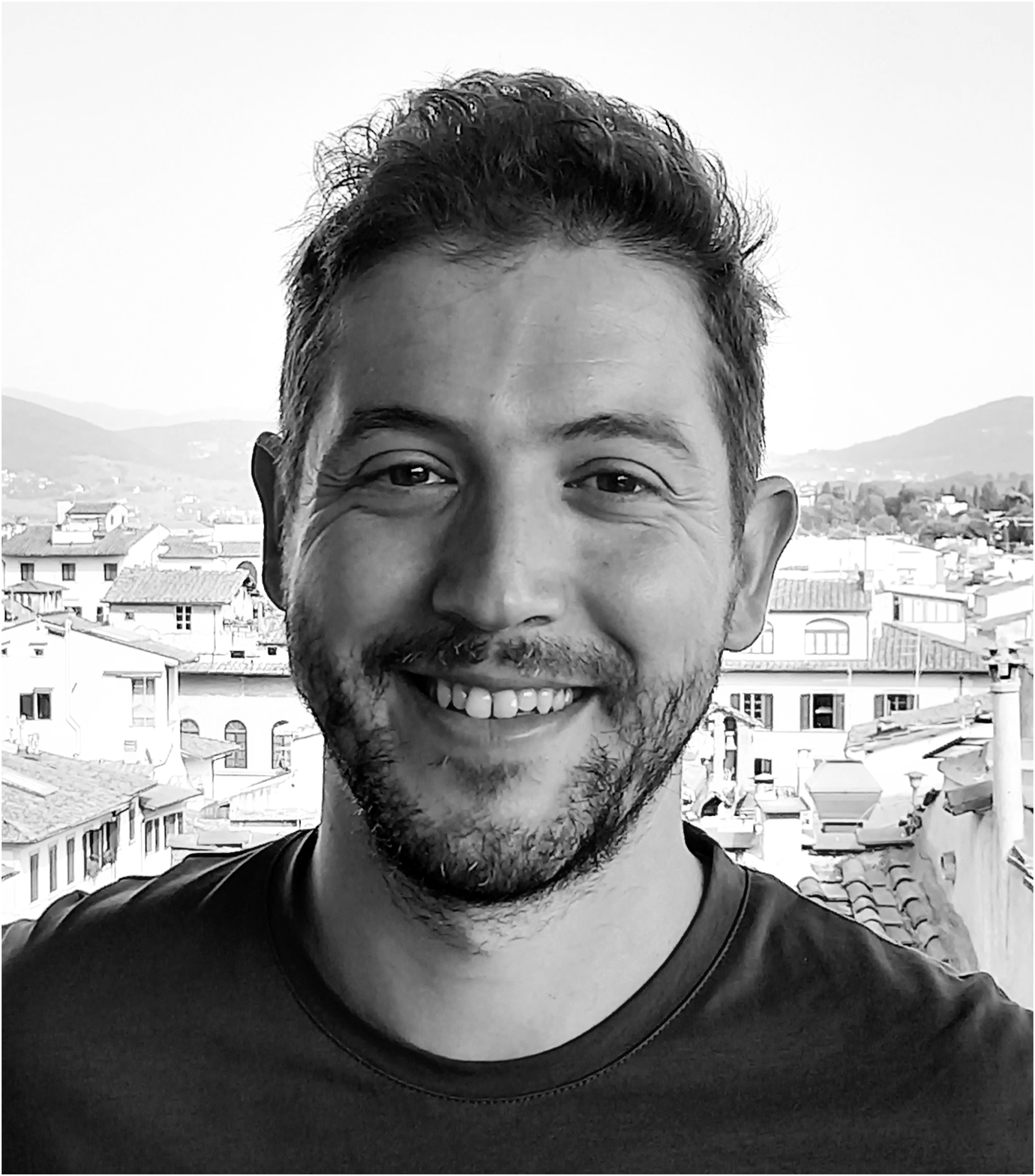}}]{Anastasios Giovanidis} (Member, IEEE) received
the Diploma degree in electrical and computer engineering
from the National Technical University of
Athens, Greece, in 2005, the Dr.-Ing. degree
from the Technical University of Berlin, Germany,
in 2010, and the Habilitation to supervise thesis (HDR) from Sorbonne University, Paris, France, in 2020.

He has held research positions with the Zuse
Institute Berlin, Germany, the National Institute
for Research in Computer Science and Automation
(Inria), France, and T\'el\'ecom ParisTech, France.
He is currently a permanent Researcher with the
French National Center for Scientific Research (CNRS), LIP6, Sorbonne
Université, Paris, France. His research interests are in mathematical modeling,
data analysis and optimization of wireless, content delivery, and social
networks.
\end{IEEEbiography}




\newpage
\appendices
\section{appendix}
\label{appendix:a}




\subsection{Proof of Lemma~\ref{lema2}}
\label{A-1}
For the proof we will first need the following.
\begin{lemma}
\label{lema1}
Let us consider $Z \in \mathbb{N},$ the vector $\rho \in \mathbb{R}_{> 0}^{Z}$ by $\rho=(\rho_1,...,\rho_Z),$ 
$\mathbf{r}=(r_1,...,r_Z) \in \mathbb{R}_{\geq 0}^{Z},$ $\mathbf{y}=(y_0,y_1,...,y_Z) \in \mathbb{R}^{Z+1}$, $K=(K_1,...,K_Z) \in \mathbb{R}^{Z}$ and a budget $B \geq 0$. In addition, consider the next linear programming problem: 
\begin{align*}
\label{minlin}
\textrm{min}_{\Tilde{\mathbf{y}}} \quad & B y_0+\sum^{Z}_{j=1} r_j  y_j, \\
\textrm{ s.t.} \quad & \rho_n y_0+y_n \geq K_n,  \forall n \in \{ 1,2,...,Z\} \tag*{\textit{[LD]}},\\
& 0 \leq y_0, \ 0 \leq y_n,  \forall n \in \{ 1,2,...,Z\}. 
\end{align*}

Suppose the users $\{i_{k} \}_{k=1,...,Z}$ are indexed by decreasing order by their $\frac{K_{i_{k}}}{\rho_{i_k}}$ and let us define $\tau$ as the maximum $\tau$ such that $\sum^{\tau}_{m=1}  \rho_{i_{m}} r_{i_{m}} \leq B.$ Then a solution $\mathbf{y}^{*}$ to~\ref{minlin} is given for all $j \in \{0,1,...,Z \}$ as:
\[
  {y^{*}_j =
  \begin{cases}
                               \frac{K_{i_{\tau+1}}}{\rho_{i_{\tau+1}}} & \text{if $j=0$}, \\
                                   K_{i_{k}}-\rho_{i_{k}} \frac{K_{i_{\tau+1}}}{\rho_{i_{\tau+1}}} & \text{if $j=i_k$ and $k \in \{1,...,\tau \}$}, \\
                                   0 & \text{if $j=i_k$ and $k \in \{\tau+1,...,Z \}$}.
  \end{cases}}
\]

\end{lemma}
\begin{proof}
Let us proceed to find a lower bound of the objective function $B y_0+\sum^{Z}_{j=1} r_j  y_j$. For this purpose let
$\mathbf{y}^{*}=(y^{*}_0,y^{*}_1,...,y^{*}_Z) \in \mathbb{R}^{Z+1}$ be a solution to~\ref{minlin}, thus it holds:

\begin{equation}
\label{eq12}
    y^*_{i_k} \geq K_{i_{k}}-\rho_{i_{k}} y^*_0,  \forall k \in \{ 1,2,...,\tau\},
\end{equation}

By multiplying inequality~(\ref{eq12}) with $r_{i_k} \geq 0$ and summing over the indices $k \in \{ 1,2,...,\tau\}$ we have:
\begin{equation}
    \label{eq122}
    \sum^{\tau}_{k=1} r_{i_k} y^*_{i_k} \geq \sum^{\tau}_{k=1} [r_{i_k}(K_{i_{k}}-\rho_{i_{k}} y^*_0)], 
\end{equation}


Note that $\sum^{\tau}_{m=1}  \rho_{i_{m}} r_{i_{m}}+\rho_{i_{\tau+1}} r_{i_{\tau+1}} > B$ by construction of $\tau$, hence:
\begin{equation}
    \label{eq14}
    r_{i_{\tau+1}}>\frac{B-\sum^{\tau}_{m=1}  \rho_{i_{m}} r_{i_{m}}}{\rho_{i_{\tau+1}}} \geq 0,
\end{equation}

On the other hand, since that $\mathbf{y}^{*}$ is a solution to~\ref{minlin}, so we have:
\begin{equation}
    \label{eq15}
    y^*_{i_{\tau+1}} \geq (K_{i_{\tau+1}}-\rho_{i_{\tau+1}} y^*_0),
\end{equation}

By hypothesis $y^*_{i_{\tau+1}} \geq 0$, thus the order of multiplying the inequality~(\ref{eq14}) and inequality~(\ref{eq15}) holds, namely:
\begin{equation}
    \label{eq16}
    r_{i_{\tau+1}} y^*_{i_{\tau+1}} \geq (K_{i_{\tau+1}}-\rho_{i_{\tau+1}} y^*_0) \frac{B-\sum^{\tau}_{m=1}  \rho_{i_{m}} r_{i_{m}}}{\rho_{i_{\tau+1}}},
\end{equation}

Observe that $r_{i_{k}} \geq 0$ and $y^*_{i_{k}} \geq 0$, for all $k \in \{\tau+2,...,Z\}$, then we have:
\begin{equation}
    \label{eq17}
    \sum^{Z}_{k=\tau+2} r_{i_{k}} y^*_{i_{k}} \geq 0,
\end{equation}

Adding up the inequality~(\ref{eq122}), inequality~(\ref{eq16}) and inequality~(\ref{eq17}), we obtain:
\begin{equation} 
\label{eq19}
    B y^*_0+\sum^{Z}_{j=1} r_j  y^*_j \geq \sum^{\tau}_{k=1} (r_{i_k}K_{i_{k}})+K_{i_{\tau+1}} \frac{B-\sum^{\tau}_{m=1}  \rho_{i_{m}} r_{i_{m}}}{\rho_{i_{\tau+1}}},
\end{equation}

Therefore, we have found a lower bound of the objective function described by the right-hand side of inequality~(\ref{eq19}). We proceed to construct a feasible point $y^0_j \in \mathcal{M}$ that reaches the lower bound described by the inequality~(\ref{eq19}) and we shall have concluded. So, we construct $y^0_j$ defined as follows:
\[
  {y^0_j =
  \begin{cases}
                               \frac{K_{i_{\tau+1}}}{\rho_{i_{\tau+1}}} & \text{if $j=0$}, \\
                                   K_{i_{k}}-\rho_{i_{k}}  \frac{K_{i_{\tau+1}}}{\rho_{i_{\tau+1}}} & \text{if $j=i_k$ and $k \in \{1,...,\tau \}$}, \\
                                   0 & \text{if $j=i_k$ and $k \in \{\tau+1,...,Z \}$}.
  \end{cases}}
\]

Let us prove that $y^0_j \in \mathcal{M}.$ Since that $\{i_{k} \}_{k=1,...,Z}$ are indexed by decreasing order by their $\frac{K_{i_{k}}}{\rho_{i_k}},$ thus $\frac{K_{i_{k}}}{\rho_{i_k}} \geq \frac{K_{i_{k+1}}}{\rho_{i_{k+1}}}$ or equivalently $
y^0_{i_k} \geq 0, \forall k \in \{ 1,2,...,\tau\}.$ Hence, $y^0_0 \geq 0$ and $y^{0}_n \geq 0,  \forall n \in \{ 1,2,...,Z\}.$ On the other hand, observe that for all $k \in \{ 1,2,...,\tau\}$:
\begin{equation}
    \rho_{i_k} y^0_0+y^0_{i_k} = \rho_{i_k}  
\frac{K_{i_{\tau+1}}}{\rho_{i_{\tau+1}}}+[K_{i_{k}}-\rho_{i_{k}} \frac{K_{i_{\tau+1}}}{\rho_{i_{\tau+1}}}] \geq K_{i_k},
\end{equation}

In turn by construction of $\tau$, we also have:
\begin{equation}
    \rho_{j} y^0_0+y^0_{j} = \rho_{j}  
\frac{K_{i_{\tau+1}}}{\rho_{i_{\tau+1}}} \geq K_{j}. \forall j \in \{\tau+1,...,Z\},
\end{equation}

Therefore, $y^0$ is a feasible point.

Finally, observe that $y^0$ reaches the lower bound, namely it matches with the right-hand side of inequality~(\ref{eq19}) because:
\begin{equation}
\begin{split}
    B y^0_0+\sum^{Z}_{j=1} [r_j  y^0_j] & = B \frac{K_{i_{\tau+1}}}{\rho_{i_{\tau+1}}}+\sum^{\tau}_{k=1} [r_{i_k}( K_{i_{k}}-\rho_{i_{k}}  \frac{K_{i_{\tau+1}}}{\rho_{i_{\tau+1}}})], \\
    & = \sum^{\tau}_{k=1} (r_{i_k}K_{i_{k}})+K_{i_{\tau+1}} \frac{B-\sum^{\tau}_{m=1}  \rho_{i_{m}} r_{i_{m}}}{\rho_{i_{\tau+1}}}.
\end{split}
\end{equation}

Hence, $y^0$ is a solution to~\ref{minlin} and we have the result.
\end{proof}

We can now proceed with the proof of Lemma~\ref{lema2}.

\begin{proof}
Let us first note that if $l \notin \mathcal{Z}$ and $l \not =i$, then by construction of $\mathcal{Z}$ we have that $\rho_l s_l=0, \forall s_l \in [0,r_l]$. Observe that to maximize the objective function $\sum_{j \in \mathcal{N}\setminus \{i \} } K_j  s_j$ of the problem~\ref{maxlin} with respect $s_l \in [0,r_l]$ is equivalent to maximize $K_l s_l$ with respect $s_l \in [0,r_l]$. Trivially, $K_l s_l$ is maximized with $s^*_l=r_l I_{K_l > 0}.$ Thus if $j \notin \mathcal{Z}$ and $j \not =i,$ then $s^*_j=r_j I_{K_j > 0}.$

Therefore, without loss of generality we can suppose that $\mathcal{Z}=\mathcal{N}\setminus \{i \},$ namely, we can suppose that the problem~\ref{maxlin} is of the form:
\begin{align*}
\label{maxlin2}
\textrm{max}_{\mathbf{s}} \quad & \sum_{j \in \mathcal{N}\setminus \{i \} } K_j  s_j, \\
\textrm{ s.t.} \quad & \sum_{n \in \mathcal{N}\setminus \{i \} }  \rho_n s_n \leq B, \tag*{\textit{[LP]}}\\
& 0 \leq s_n \leq r_n,  \forall n \in \mathcal{N}\setminus \{i \}. 
\end{align*}
With $\rho_n \not =0, \forall n \in \mathcal{N}\setminus \{i \}.$

Let us observe that the construction of the Lagrange dual of the standard form of the linear problem~\ref{maxlin2} (see \cite{III0} pp. 223-227) is:
\begin{align*}
\label{minlin2}
\textrm{min}_{\mathbf{y}} \quad & B y_0+\sum_{j \in  \mathcal{N}\setminus \{i \}} r_j  y_j, \\
\textrm{ s.t.} \quad & \rho_n y_0+y_n \geq K_n,  \forall n \in \mathcal{N}\setminus \{i \} \tag*{\textit{[LD]}},\\
& 0 \leq y_0, \ 0 \leq y_n,  \forall n \in \mathcal{N}\setminus \{i \}. 
\end{align*}
With $\mathbf{y}=(y_0,y_1,...,y_{i-1},y_{i+1},...,y_{N}) \in \mathbb{R}^{N}$ the variables in the dual formulation.

By Lemma~\ref{lema1} a solution $\mathbf{y}^*$ of the problem~\ref{minlin2} is:
\[
  {y^{*}_j =
  \begin{cases}
                               \frac{K_{i_{\tau+1}}}{\rho_{i_{\tau+1}}} & \text{if $j=0$}, \\
                                   K_{i_{k}}-\rho_{i_{k}} \frac{K_{i_{\tau+1}}}{\rho_{i_{\tau+1}}} & \text{if $j=i_k$ and $k \in \{1,...,\tau \}$}, \\
                                   0 & \text{if $j=i_k$ and $k \in \{\tau+1,...,Z \}$}.
  \end{cases}}
\]

Let us define $\Tilde{s}$ as follows:
\[
  \Tilde{s}_j =
  \begin{cases}
                                   r_{j} & \text{if $j=i_k \in \mathcal{Z}$ and $k \leq \tau$}, \\
                                   \frac{B-\sum^{\tau}_{k=1} \rho_{i_k} r_{i_k}}{\rho_{i_{\tau+1}}}
                                   & \text{if $j=i_k \in \mathcal{Z}$ and $k=\tau+1$},\\
                                   0 & \text{if $j=i_k \in \mathcal{Z}$ and $k \geq \tau+2$},
  \end{cases}
\]

Observe that $\Tilde{s}$ is a feasible point of the problem~\ref{maxlin2} since $0 \leq \Tilde{s}_n \leq r_n$ by construction, and $\sum_{n \in \mathcal{N}\setminus \{i \} }  \rho_n \Tilde{s}_n=\sum^{\tau}_{k=1} (\rho_{i_k} r_{i_k})+\frac{B-\sum^{\tau}_{k=1} \rho_{i_k} r_{i_k}}{\rho_{i_{\tau+1}}} \rho_{i_{\tau+1}} \leq B.$

Note that the optimal of the problem~\ref{minlin2} matches with the evaluation of the feasible point $\Tilde{s}$ in the objective function of the problem~\ref{maxlin2} as follows:

\begin{equation}
    \label{22}
    \begin{split}
    \sum_{j \in \mathcal{N}\setminus \{i \} } K_j  \Tilde{s}_j &=  \sum^{\tau}_{k=1} (r_{i_k} K_{i_{k}})+K_{i_{\tau+1}} \frac{B-\sum^{\tau}_{k=1}  \rho_{i_{k}} r_{i_{k}}}{\rho_{i_{\tau+1}}},\\
    &= B \frac{K_{i_{\tau+1}}}{\rho_{i_{\tau+1}}}+\sum^{\tau}_{k=1} [r_{i_k} (K_{i_{k}}-\rho_{i_k} \frac{K_{i_{\tau+1}}}{\rho_{i_{\tau+1}}})],\\
    &= By^*_0+\sum_{j \in \mathcal{N}\setminus \{i \} } r_j  y^*_j.
    \end{split}
\end{equation}

On the other hand, notice that by \cite{III0} pp.223-227, the duality gap of the problem~\ref{maxlin2} and the problem~\ref{minlin2} is zero. Hence, a solution $s^*$ of the problem~\ref{maxlin2} holds $\sum_{j \in \mathcal{N}\setminus \{i \} } K_j  s^*_j=By^*_0+\sum_{j \in \mathcal{N}\setminus \{i \} } r_j  y^*_j=\sum_{j \in \mathcal{N}\setminus \{i \} } K_j  \Tilde{s}_j,$ therefore $\Tilde{s}$ is solution of the problem~\ref{maxlin2} and we have the desired result.
\end{proof}

\subsection{Proof of Theorem~\ref{teo1}}
\label{A-2}
%
\begin{proof}
Let us first note that by Lemma~\ref{lema2}, we have that the solution to the problem $\mathbf{s}^{(t)} \in \argmax_{\mathbf{s}^{(t)} \in \mathcal{M}} \langle \nabla U(\mathbf{a}^{(t)}), \mathbf{s}^{(t)} \rangle$ is given by:
\[
  \mathbf{s}_j^{(t)} =
  \begin{cases}
                                r_j I_{\nabla_j U(\mathbf{a}^{(t)}) > 0} & \text{if $j \notin \mathcal{Z}$ and $j \not =i$}, \\
                                   r_{j} & \text{if $j=i_k \in \mathcal{Z}$ and $k \leq \tau$}, \\
                                   \frac{B-\sum^{\tau}_{k=1} c_{i_k} \lambda ^ {(i_{k})} r_{i_k}}{c_{i_{\tau+1}} \lambda ^ {(i_{\tau+1})}}
                                   & \text{if $j=i_k \in \mathcal{Z}$ and $k=\tau+1$},\\
                                   0 & \text{if $j=i_k \in \mathcal{Z}$ and $k \geq \tau+2$},
  \end{cases}
\]
With $\mathcal{Z}=\{k \in \mathcal{N}\setminus \{i \} | c_k \lambda ^{(k)} \not =0\}$, $\{i_{k} \}_{k=1,...,|\mathcal{Z}|}$ are indexed by decreasing order by their $\frac{\nabla_{i_{k}} U(\mathbf{a}^{(t)})}{c_{i_k} \lambda ^{(i_k)}}$ and $\tau$ is defined as the maximum $\tau$ such that $\sum^{\tau}_{m=1}  c_{i_m} \lambda ^{(i_m)} r_{i_{m}} \leq B.$ So, Algorithm~\ref{algo3} solves the sub-problem $\argmax_{\mathbf{s}^{(t)} \in \mathcal{M}} \langle \nabla U(\mathbf{a}^{(t)}), \mathbf{s}^{(t)} \rangle$ and uses it in the Frank-Wolfe algorithm described in Algorithm~\ref{algo2}.

Therefore~(\ref{ceq25}) and~(\ref{ceq26}) are a direct implication by \cite[Theorem 1 and 2]{III2} since $\mathcal{M}$ is compact and convex. 

Let us now prove that the computational complexity in memory of Algorithm~\ref{algo2} is of order $\mathcal{O}(max(N-1,D))$ respectively. Note that the vectors $\mathbf{a}^{(0)},\omega^{(0)},\nabla U(\mathbf{a}^{(0)}),\mathbf{s}^{(0)},\mathbf{d}_0,\mathbf{g}_0$ and the step size $\gamma_0$ can be stored in memory only once and updated at each step in the Algorithm~\ref{algo2}, so the computational complexity in memory will be $\mathcal{O}(N-1)+\mathcal{O}(D)=\mathcal{O}(max(N-1,D))$ since that the average impression ratios $\{p^{(j)}_{n} \not = 0 | (n,j) \in \mathcal{N} \times \mathcal{N}\}$ are stored in memory efficiently, and that timsort can be used in Algorithm~\ref{algo3} with a worst-computational complexity in memory of order $\mathcal{O}(N-1).$

On the other hand, observe that by construction of~\ref{[BPO-G]} we have that the vectors $\mathbf{d}_t,\mathbf{g}_t$ and $\mathbf{a}^{(t+1)}$ take at most $\mathcal{O}(N-1)$ computations to update, however $\omega^{(t)}$ and $\nabla U(\mathbf{a}^{(t)})$ take at most $\mathcal{O}(D)$ computations to update. In the Algorithm~\ref{algo3}, notice that $\mathcal{Z}$ takes at most $\mathcal{O}(N-1)$ computations to obtain, however timsort takes in an average case and worst case $\mathcal{O}((N-1) log(N-1))$ computations by \cite{III6}, and in a best case $\mathcal{O}(N-1)$ computations by \cite{III7}. Note that $\mathbf{s}$ takes $\mathcal{O}(N-1)$ computations to get. Therefore, Algorithm~\ref{algo2} has an average- and worst- computational complexity per step in time of order $\mathcal{O}((N-1) log(N-1))+\mathcal{O}(D)=\mathcal{O}(max((N-1) log(N-1),D)),$ and a best-computational complexity per step in time of order $\mathcal{O}(N-1)+\mathcal{O}(D)=\mathcal{O}(max(N-1,D)).$
\end{proof}

\subsection{Iterative algorithm for multi-instances}
\label{a:3}

A low-complexity fast algorithm similar to Algorithm~\ref{algo2} in $L Q (N-1)$ variables for the set $D=\{(l,q,n,j) \in \mathcal{L} \times \mathcal{Q} \times \mathcal{N} \times \mathcal{N} | p^{(j)}_{l,q,n} \not = 0\}$ can be obtained by updating:
\begin{equation}
    \omega^{(t)}_{l,q,j} \gets \sum_{n \in \mathcal{N}\setminus \{j \}} \zeta_{l,q} a_{l,q,n} p^{(j)}_{l,q,n}, 
\end{equation}
\begin{equation}
    \nabla_{l,q,k} U(\mathbf{a}^{(t)}) \gets \sum_{j \in \mathcal{N}\setminus \{i,k \}} \sigma_l \zeta_{l,q} p^{(j)}_{l,q,k} U'_j(\omega^{(t)}_{l,q,j}).
\end{equation}

Hence, the solution vector of the optimization problem~\ref{[M-BPO]} 
can be found by a fast algorithm with space complexity in memory of order $\mathcal{O}(max(L Q (N-1),D))$ and average-computational complexity in time of order $\mathcal{O}(max(L Q (N-1) \log(L Q (N-1)),D))$ per iteration. The above algorithm is an $\epsilon$-approximate solution to the problem in the Primal after $\mathcal{O}(1/\epsilon)$ steps. In practice, the previous algorithms are fast in sparse environments where $D \ll L Q N^{2}.$


\ifCLASSOPTIONcompsoc

\end{document}